\documentclass[a4paper,12pt]{article}
\pdfoutput=1 

\usepackage{jcappub} 
\usepackage{subfigure}
\usepackage[T1]{fontenc} 
\usepackage[dvipsnames*,svgnames]{xcolor}
\definecolor{maincolor}{rgb}{0,1,0.5}
\definecolor{thecolor}{rgb}{0.1,0.5,1}

\title{\LARGE\boldmath Searching for Secluded Dark Matter with H.E.S.S., Fermi-LAT, and Planck}

\author[a]{Stefano Profumo,}
\author[b,c]{Farinaldo S. Queiroz,}
\author[d,e,f]{Joseph Silk,}
\author[b,g]{Clarissa Siqueira}

\affiliation[a]{Department of Physics and Santa Cruz Institute for Particle Physics,\\ University of California, Santa Cruz, CA 95064, USA}
\affiliation[b]{Max-Planck-Institut f\"ur Kernphysik, Saupfercheckweg 1, 69117 Heidelberg,
Germany}
\affiliation[c]{International Institute of Physics, Federal University of Rio Grande do Norte, Campus Universit\'ario,
Lagoa Nova, Natal-RN 59078-970, Brazil}
\affiliation[d]{Institut d'Astrophysique, CNRS, UPMC, Paris 75014, France}
\affiliation[e]{Department of Physics and Astronomy, The Johns Hopkins University Homewood Campus, Baltimore, MD 21218, USA}
\affiliation[f]{BIPAC, Department of Physics, University of Oxford, Keble Road, Oxford OX1 3RH, UK}
\affiliation[g]{Departamento de F\'isica, Universidade Federal da Para\'iba, Caixa Postal 5008, 58051-970, Jo\~ao Pessoa, PB, Brasil}

\emailAdd{profumo@ucsc.edu}
\emailAdd{queiroz@mpi-hd.mpg.de}
\emailAdd{joseph.silk@physics.ox.ac.uk}
\emailAdd{clarissa@mpi-hd.mpg.de}

\abstract{
Short-lived mediators are often used to describe dark matter interactions with Standard Model particles. When the dark matter mass is heavier than the mass of the mediator, it may self-annihilate into short-lived mediators, and in some cases this might be the dominant annihilation channel. This scenario is known as secluded dark matter. We use Fermi-LAT observations of dwarf spheroidal galaxies, H.E.S.S. data from the  Galactic center, and Planck measurements of the Cosmic Microwave Background to constrain secluded dark matter. We explore the interplay between these experiments and we assess the impact of the mediator mass on our bounds, an often overlooked yet very important point.  In particular, we exclude pair -annihilation cross-sections greater or on the order of $\sigma v \sim 4 \times 10^{-27} {\rm cm^3/s}$ for dark matter masses around $10$~GeV and greater or on the order of $\sigma v \sim \times 10^{-25} {\rm cm^3/s}$ for dark matter masses around a TeV. Our findings supersede previous constraints which use Fermi-LAT data, and constitute the first limits on secluded dark sectors using the H.E.S.S. telescope.  We also show that one can fit TeV gamma-ray observations from H.E.S.S. with secluded dark matter annihilations, with the mediator mass impacting the best-fit dark matter particle mass. Our findings indicate that any assessment of secluded dark sectors in the context of indirect detection significantly depends on the choice of the mediator mass.
}

\begin{document}
\maketitle
\flushbottom

\section{Introduction}
\label{sec:intro}

The nature of dark matter is one of the most exciting and important open questions in science \cite{Bertone:2004pz,Strigari:2013iaa,Profumo:2013yn,Bertone:2016nfn,Queiroz:2016sxf}. The Standard Model (SM) does not have a fundamental particle capable of explaining the cold dark matter in our universe, which constitutes 27\% of the entire energy-budget. 
 Among proposed dark matter candidates, Weakly Interacting Massive Particles (WIMPs) stand out as uniquely well motivated, since via the thermal freeze-out, they could naturally explain the observed dark matter abundance \cite{Arcadi:2017kky}.\\
 
Among different detection techniques, indirect dark matter detection, the detection of the debris of dark matter annihilations or decays, has the remarkable advantage of connecting particle physics and astrophysics: a potential dark matter signal would pinpoint the dark matter mass and annihilation cross-section, provided certain astrophysical inputs \cite{Bringmann:2012ez}.\\

Typically, indirect searches for dark matter are focused on final states belonging to the SM spectrum such as quarks, leptons and gauge bosons \cite{Bringmann:2011ye,Hooper:2012sr,Ibarra:2013zia,Bergstrom:2013jra,Lin:2014vja,Bringmann:2014lpa,Gonzalez-Morales:2014eaa,DiMauro:2015jxa,Buckley:2015doa,Giesen:2015ufa,Lu:2015pta,Cavasonza:2016qem,Belotsky:2016tja,Profumo:2016idl,Huang:2016tfo,Caputo:2016ryl,Jin:2017iwg}. However, it is also possible that WIMPs annihilate mostly into short-lived particles which later decay into SM particles \cite{Mardon:2009rc,Cerdeno:2015ega,Okawa:2016wrr,Karwin:2016tsw,Kim:2016csm,Batell:2017rol,Campos:2017odj,Arcadi:2017vis}. This setup is very common for dark matter models in the context of the $Z^\prime$ portal when the dark matter is heavier than the $Z^\prime$ \cite{Pospelov:2007mp,Mardon:2009rc,Kang:2010mh,Murase:2012xs,Alves:2013tqa,Martinez:2014ova,DEramo:2016gos,Celis:2016ayl,DeRomeri:2017oxa}, in vector dark matter models \cite{Ko:2014gha}, also in models where the dark matter is entirely confined to the dark sector but annihilates to particles that subsequently decay to SM products \cite{Kim:2016csm,Ardid:2017lry} as well as in studies where bounds states are formed \cite{Cirelli:2016rnw} (see \cite{Fortes:2015qka,Fortes:2017kca} for other scenarios where secluded dark matter arises). \\

Motivated by the presence of such ``secluded annihilations'' in several models, numerous studies have been performed in the context of indirect dark matter detection \cite{Pospelov:2008jd,Batell:2009zp,Abdullah:2014lla,Martin:2014sxa,Dutta:2015ysa,Rajaraman:2015xka,Elor:2015tva,Escudero:2017yia}. The present study differs from previous work in several key
aspects:\\

{\bf (i)} We do not focus on one specific dataset, and we consistently include Fermi-LAT, H.E.S.S. and Planck data to draw our conclusions;

{\bf (ii)} We use seven years of Fermi-LAT data in the direction of dozens of dwarf spheroidal galaxies (dSphs) with the updated instrument response function (PASS8) \cite{Ackermann:2015zua};

{\bf (iii)} We make use of 10 years of H.E.S.S. data, equivalent to 254h live-time exposure, on the Galactic center with the latitude region $|b| < 3^\circ$ removed \cite{Abdallah:2016ygi}; 

{\bf (iv)} We account for Planck CMB data, as such data is rather sensitive to any injection of electromagnetically interacting particles capable of producing appreciable ionization and heat, thus severely constraining dark matter annihilations at early times \cite{Slatyer:2015jla,Slatyer:2015kla};

{\bf (v)} We critically and quantitatively assess the impact of the mediator mass on our bounds. We show that the mass of the short-lived mediator does matter in the derivation of precise limits on the dark matter annihilation cross-section and in the determination of best-fit regions of parameter space. In the context of indirect dark matter detection, these mediators are typically assumed to simply be sufficiently heavier than the  SM particle pairs produced in their decay modes. However, if one departs from this particular assumption, the experimental limits change significantly;

{\bf (vi)} We discuss the possibility of explaining the TeV gamma-ray excess observed by the H.E.S.S. telescope in the Galactic center \cite{Aharonian:2009zk,Abramowski:2016mir} with secluded dark matter annihilations.\\

Our study is structured as follows: In {\it section 2}, we lay out the setup that we focus on; in {\it section 3} we discuss how we derive the production of gamma-rays and ionization induced by dark matter annihilation; in {\it section 4} we describe the dataset we use; in {\it section 5} we present our limits, {\it section 6} we address the H.E.S.S. TeV excess in terms of secluded dark matter and finally, in {\it section 7}, we draw our conclusions.

\section{Secluded Dark Matter}

\begin{figure}[!h]
\centering
\includegraphics[width=0.3\columnwidth]{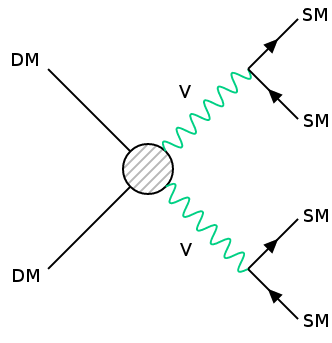}
\caption{Feynman diagrams for dark matter annihilation into short-lived states that decay into SM particles.}
\label{fig:annihilation}
\end{figure}

Secluded dark matter refers to a scenario where the dark matter particle annihilates preferentially into particles that do not belong to the SM particle content. 
Dark matter particles that annihilate directly into SM particles can be severely constrained by data since the same Feynman diagram that typically leads to dark matter annihilations is also responsible for dark matter-nucleon scattering, the subject of intense and highly constraining experimental searches \cite{Akerib:2015rjg,Amole:2016pye,Hehn:2016nll,Aalbers:2016jon,Agnese:2016cpb,Aprile:2016swn,Fu:2016ega,Akerib:2016lao,Cui:2017nnn,Aprile:2017ngb,Aprile:2017yea,Fatemighomi:2016ree}. There are ways to break this direct relation between annihilation and scattering such as via co-annihilations \cite{Edsjo:1997bg}, multi-mediators \cite{Duerr:2016tmh}, or velocity-dependent suppressions, among others \cite{DEramo:2010keq,Belanger:2012vp,Ko:2014loa,Cai:2015zza,Arcadi:2017vis}. However, one interesting possibility that does not require the existence of extra fields is to simply enforce the dark matter mass to be heavier than the mediator's mass. In this case, if the coupling between the dark matter particle and the mediator is sufficiently large, the main annihilation channel can be into a pair of short-lived mediators, which constitutes a paradigmatic secluded dark matter setup \cite{Pospelov:2007mp,Essig:2009jx,Alves:2015pea,Ducu:2015fda,Alves:2015mua,Okada:2016tci,Duerr:2016tmh,Jacques:2016dqz}. Another viable secluded dark matter scenario occurs when the dark matter particle belongs to either a $SU(N)$ group \cite{Karam:2015jta,Karam:2016rsz,Arcadi:2016kmk}, or a hidden valley setup \cite{Strassler:2006im,Cassel:2009pu}, or interacts with SM particles via the Chern-Simons portal \cite{Arcadi:2017jqd}. Either way, dark matter annihilation into short-lived mediators occur, and can be the most important final state. Therefore,  one can constrain the parameter space of such secluded dark matter models by searching for the byproducts of this $2 \rightarrow 4$ annihilation processes (see Fig.\ref{fig:annihilation}).

There are several possible annihilation final states based on Fig.\ref{fig:annihilation}. In order to cover many possibilities in this work, we will investigate the cases which the short-lived state decays either into leptons or into quarks, and also vary the mediator mass to assess how relevant its mass is in the derivation of the bounds on the dark matter annihilation cross section. We start first describing how we compute the dark matter signal.

\section{Dark Matter Annihilation}
\label{sec:dmann}

\subsection{Gamma-rays}
The presence of dark matter in our neighborhood motivates us to search for signs of dark matter annihilations in regions where the dark matter density is expected to be very high, such as the Galactic Center and dSphs. The differential flux for dark matter annihilation in any angular direction is

\begin{equation}
\label{eq:flux}
\frac{d\Phi_\gamma(\Delta \Omega)}{dE}(E_\gamma)=\frac{1}{4\pi}\frac{\sigma v}{2 M_{DM}^2}\frac{dN_\gamma}{dE_{\gamma}}\cdot J_{ann}
\end{equation}where $J_{ann}$ is the astrophysical J-factor with, 
\begin{equation}
   J_{ann} \; =\; \int_{\Delta \Omega} d\Omega \int \rho_{DM}^2 (s) ds\;,
 \label{eq:Jfac_def}
\end{equation} where $s\; =\; s(\theta )$ is a line-of-sight variable and $\theta$ is the angle between the line of sight and the line to the center of the dark matter halo, $\sigma v$ is annihilation cross section today, i.e. non-relativistic, $M_{DM}$ the dark matter mass, and $\rho_{DM}$ the dark matter density profile which we will assume to be a Navarro-Frenk-While (NFW) profile \cite{Navarro:2008kc} where,

\begin{equation}
   \rho_{DM} (r) \; =\; \frac{\rho_s}{(r/r_s) (1+ r/r_s)^2}\;,
 \label{eq:DM_profile}
\end{equation}where $r_s$ and $\rho_s$ are the scale radius and the characteristic density respectively. The integral in Eq.\eqref{eq:Jfac_def} is computed over the line of sight. \\

We  adopt a NFW profile because it is the same profile used by the Fermi-LAT collaboration. Different profiles will induce an overall change in our limits by a constant factor depending on whether a steeper \cite{Graham:2005xx} or more core-like profile is adopted \cite{Burkert:1995yz,Salucci:2000ps}.\\

The key quantity in the derivation of our limits is the energy spectrum because we are focused on $2\rightarrow 4$ annihilations. The shape of the energy spectrum will allow us to understand the results we present below. We use the PPPC4DM code to compute the energy spectra when the V decays into charged leptons \cite{Cirelli:2010xx}. We emphasize that PPPC4DM by default assumes the mediator to be just slightly heavier than its decay products. In order to derive bounds for other hadronic channels and assess the importance of the mediator mass, we generate the energy spectra with Pythia 8 \cite{Sjostrand:2014zea}. \\

\begin{figure}[!h]
\centering
\includegraphics[width=0.49\columnwidth]{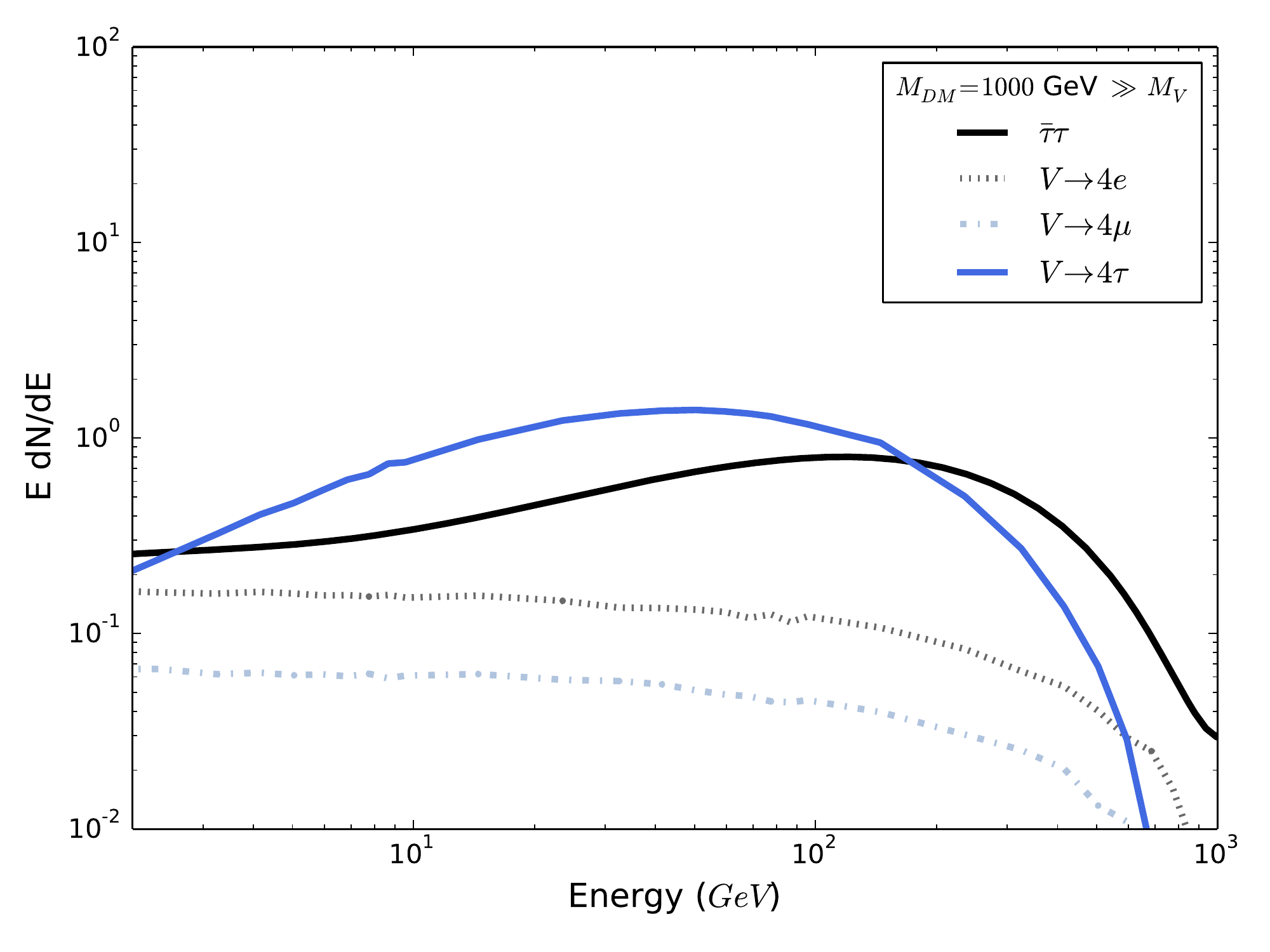}
\includegraphics[width=0.49\columnwidth]{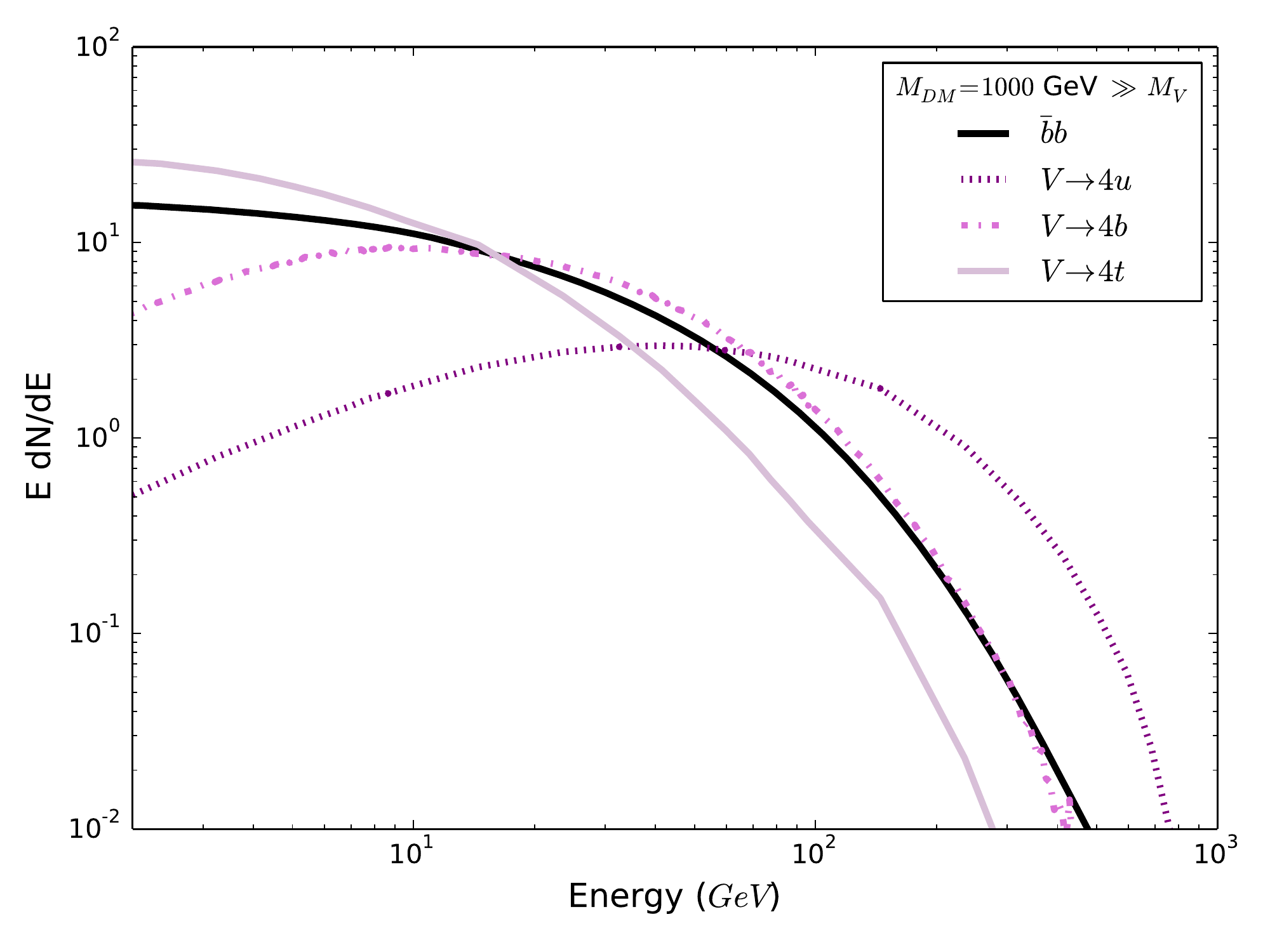}
\caption{Energy spectrum for a $1$~TeV dark matter annihilating into VV, with the V mass much larger than the final state's. Left-panel: V decays either into $ee$ or $\mu \bar{\mu}$ or $\tau \bar{\tau}$. Right-panel: V decays either into $u\bar{u}$ or $b \bar{b}$ or $t\bar{t}$. Moreover, we superimpose the direct annihilation channel ${\rm DM\, DM}\rightarrow \tau \bar{\tau}$ on the left-panel, and ${\rm DM\, DM} \rightarrow b\bar{b}$ on the right-panel for comparison.}
\label{fig1}
\end{figure}

In fig.\ref{fig1} we show the energy spectra of a $1$~TeV dark matter particle annihilating into short-lived mediators with these decaying into charged leptons (left-panel) or hadrons (right-panel). Thus far, we are assuming the dark matter particle to be much heavier than the mediator, i.e. $M_{DM} \gg M_V$.\\

By comparing the energy spectra for the dark matter annihilations into $2 \tau$ and $4\tau$ we notice that the energy spectrum is harder for the latter at lower energies between $1-100$~GeV. This occurs for the annihilations into $4b$ only during a smaller energy region, $E=20-100$~GeV, and it is much less prominent. In these figures, we take the mediator mass to be much heavier than its decay modes.\\

\begin{figure}[!h]
\centering
\includegraphics[width=0.49\columnwidth]{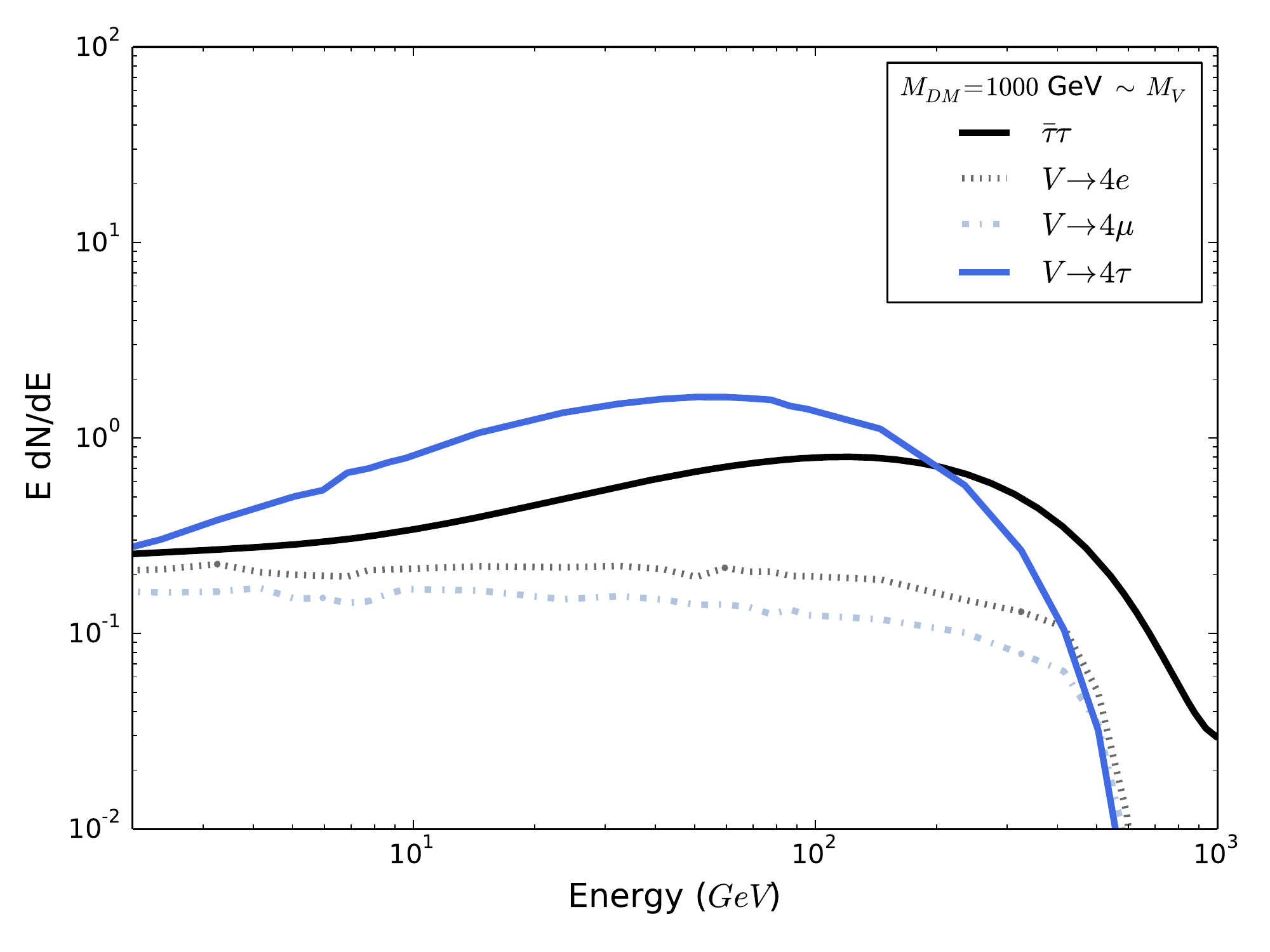}
\includegraphics[width=0.49\columnwidth]{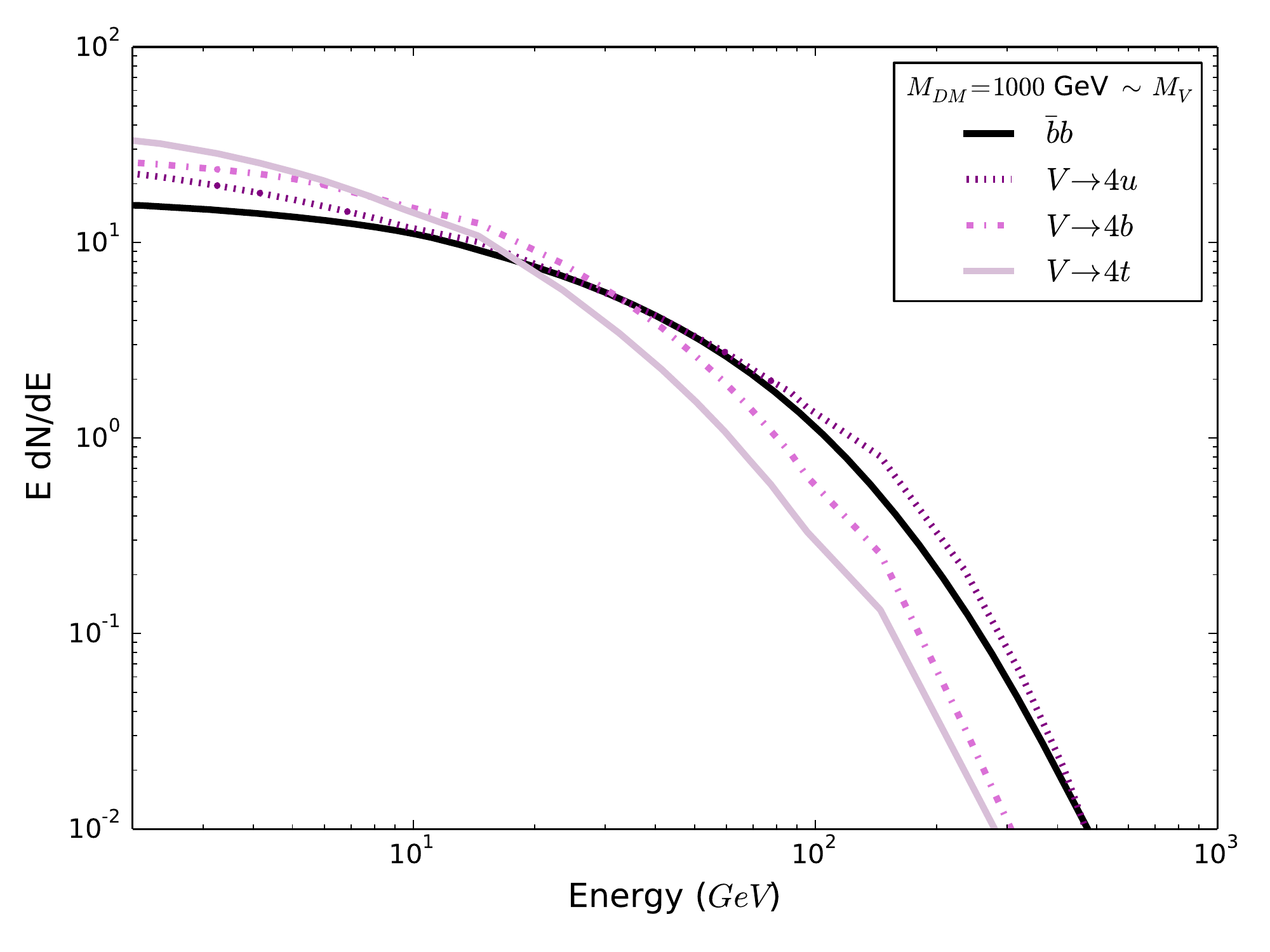}
\caption{Energy spectrum for a $1$~TeV dark matter annihilating into $VV$, with $V$ being nearly degenerate in mass. Left-panel: $V$ decays either into $ee$ or $\mu \bar{\mu}$ or $\tau \bar{\tau}$. Right-panel: $V$ decays either into $u\bar{u}$ or $b \bar{b}$ or $t\bar{t}$. Moreover, we superimpose the direct annihilation channel ${\rm DM\, DM} \rightarrow \tau \bar{\tau}$ on the left-panel, and ${\rm DM\, DM} \rightarrow b\bar{b}$ on the right-panel for comparison.}
\label{fig2}
\end{figure}

We will repeat the same procedure but take the mediator mass to be comparable to the dark matter mass. The result is displayed in Fig.\ref{fig2}. For the leptonic case, the difference in the overall energy spectrum is mild, but for the case which the mediator decays into hadrons, the difference is clearly visible, making the annihilation into $4b$ mimic the one into $2b$. \\

\begin{figure}[!h]
\centering
\includegraphics[width=0.49\columnwidth]{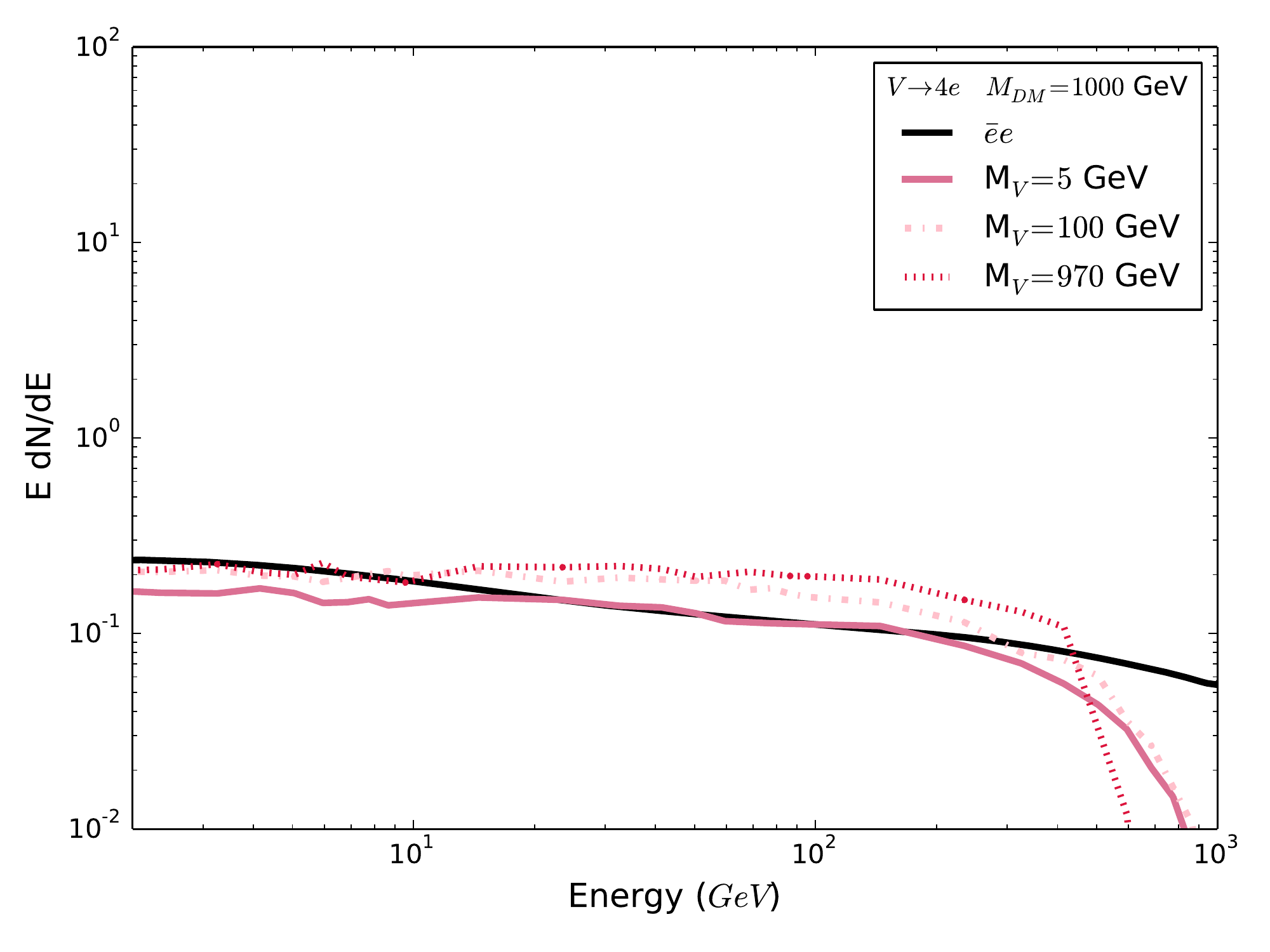}
\includegraphics[width=0.49\columnwidth]{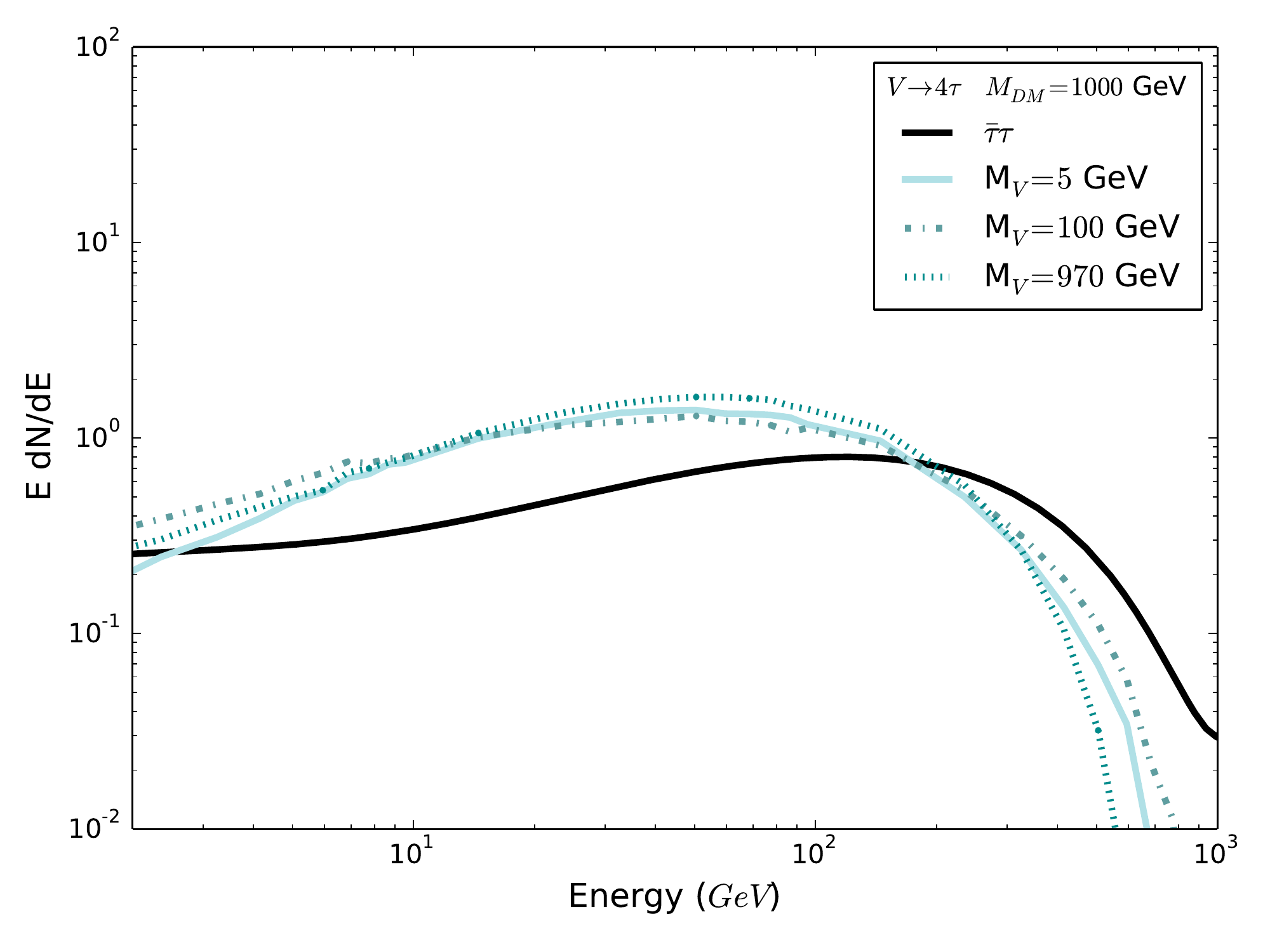}
\caption{Left-panel: energy spectrum for a $1$~TeV dark matter annihilating into $4e$ for $M_V=5, 100, 970$~GeV; Right-panel: energy spectrum for a $1$~TeV dark matter annihilating into $4\tau$ for $M_V=5, 100, 970$~GeV. We superimposed the direct annihilation channel ${\rm DM\, DM} \rightarrow e \bar{e}$ on the left-panel, and ${\rm DM\, DM} \rightarrow \tau\bar{\tau}$ on the right-panel for comparison.}
\label{fig3}
\end{figure}

\begin{figure}[!h]
\centering
\includegraphics[width=0.49\columnwidth]{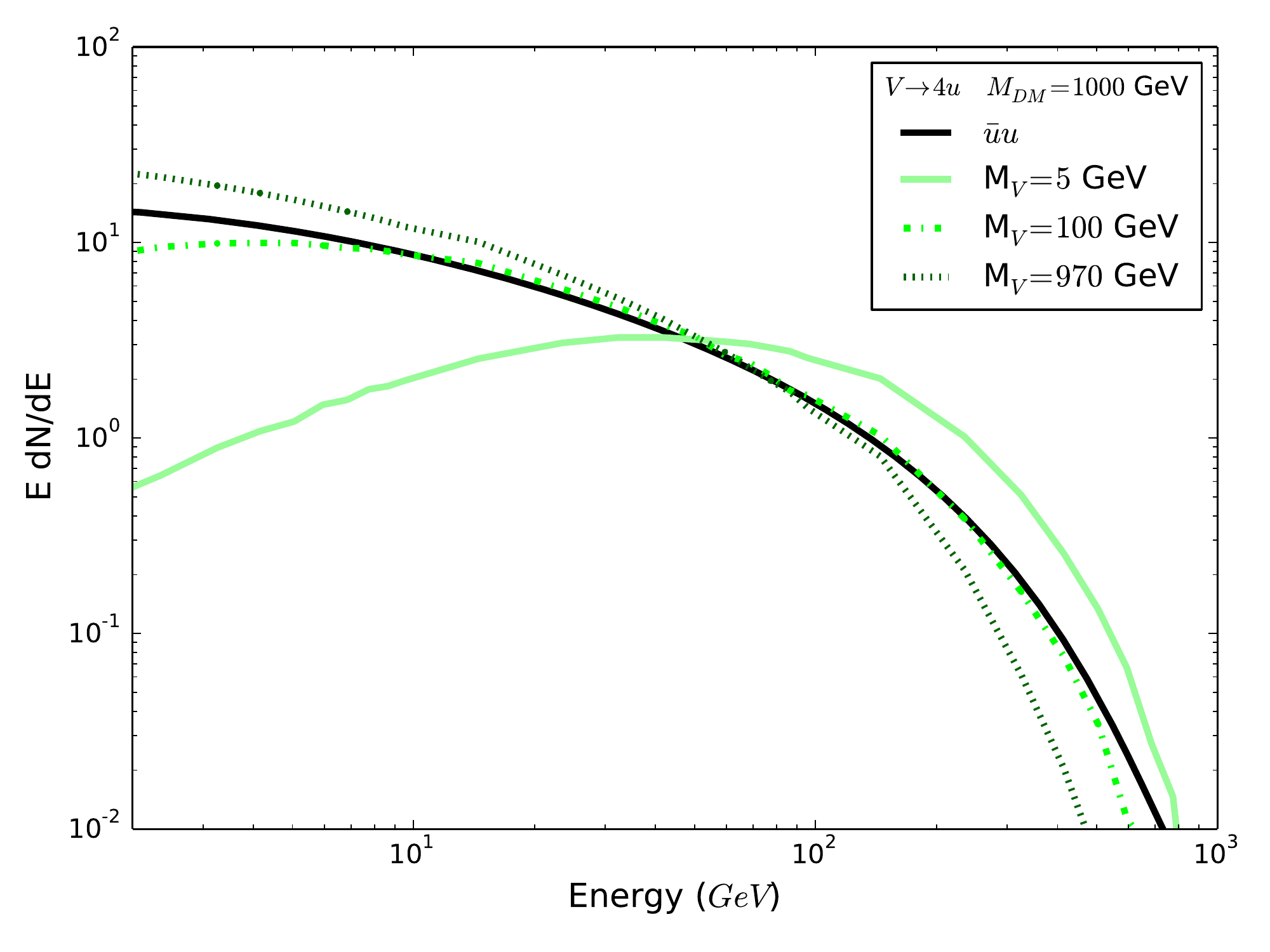}
\includegraphics[width=0.49\columnwidth]{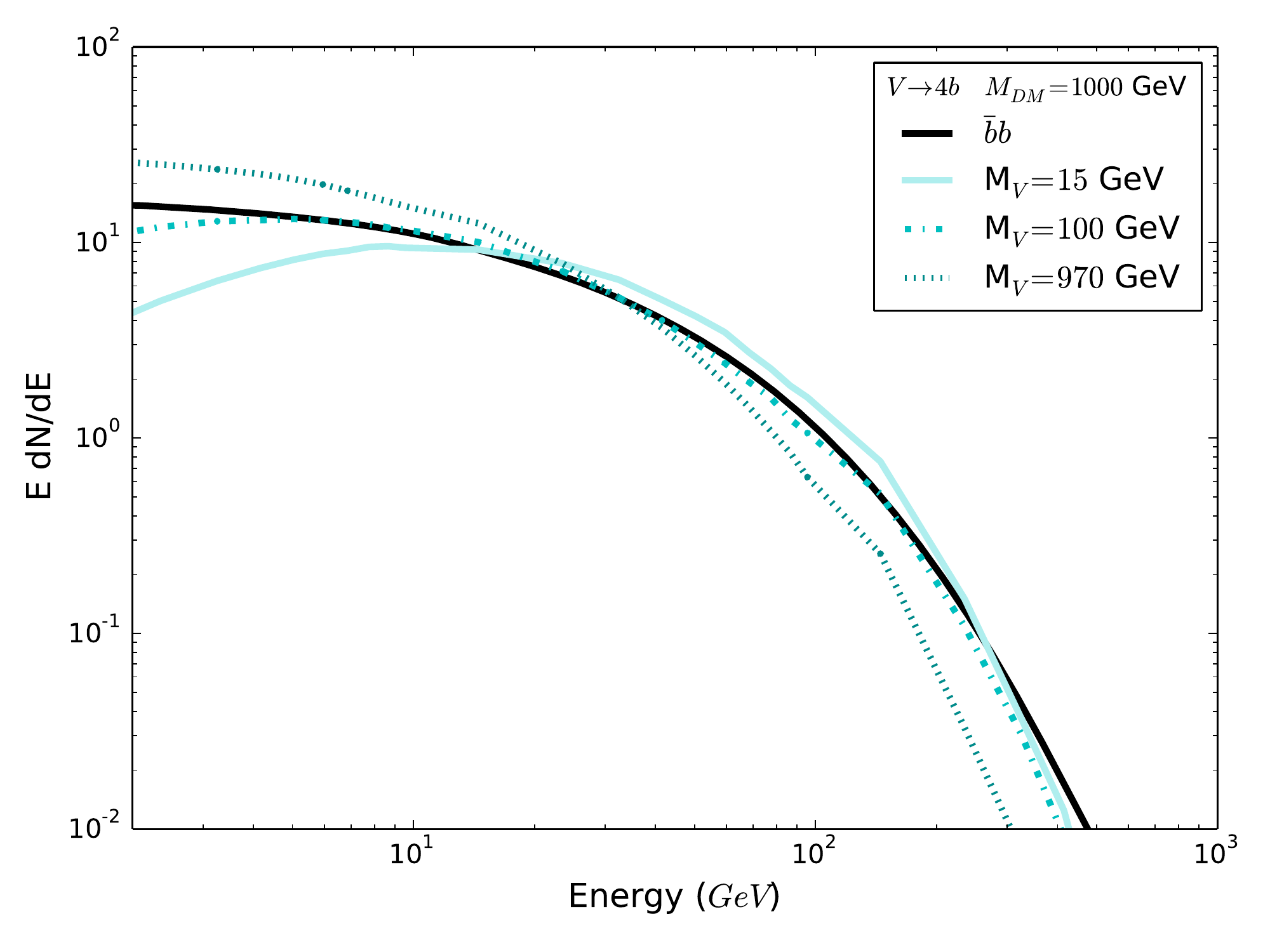}
\caption{Left-panel: energy spectrum for a $1$~TeV dark matter annihilating into $4u$ for $M_V=5, 100, 970$~GeV; Right-panel: energy spectrum for a $1$~TeV dark matter annihilating into $4b$ for $M_V=5, 100, 970$~GeV. We superimposed the direct annihilation channel ${\rm DM\, DM} \rightarrow u \bar{u}$ on the left-panel, and ${\rm DM\, DM} \rightarrow b\bar{b}$ on the right-panel for comparison.}
\label{fig4}
\end{figure}

In order to have a clearer picture of the relevance of the mediator mass in our computation, in Fig.\ref{fig3}, we derive the energy spectra for two leptonic decay modes of the mediator, $e\bar{e}$ on the left-panel and $\tau \bar{\tau}$ on the right-panel, and vary the mediator mass, while keeping $M_{DM}=1$~TeV. Notice that the mediator yields very mild differences in the energy spectra. We will see further that this does not hold true for annihilations into $4e$ and $4\mu$. Moreover, when the $V$ particle decays hadronically the mediator mass is very important. Notice in Fig.\ref{fig4} that the assumption made on the mediator mass can significantly change the energy spectra, especially towards lower energies. Therefore, any assessment of secluded dark sectors in the context of indirect detection cannot be regarded as model-independent since they rely significantly on choice for the mediator mass. \\

Now we have the gamma-ray yield for different dark matter annihilation setups, we discuss dark matter signals in the CMB.

\subsection{Cosmic Microwave Background}

In the context of thermal freeze out, dark matter annihilations are efficient in the  dark ages,  $z \leq 1200$. However, such energy injection produced by dark matter annihilations heats and ionizes the photon-baryon plasma, perturbing the ionization history precisely measured  by Planck \cite{Padmanabhan:2005es}. Such dark matter annihilations, while  not necessarily capable of changing the redshift of recombination, may alter the residual ionization after recombination which is subject to tight constraints.\\

These constraints are based on the following procedure: The energy density injection rate is determined by the product of the number density of pairs of DM particles which is $n_{DM}/2$ for Majorana particles, the
annihilation probability per time unit $P_{ann}=  \langle\sigma v \rangle n_{DM},$ and the energy released per dark matter annihilation $E_{ann}= 2 M_{DM}$. Therefore, we obtain

\begin{equation}
\frac{dE}{dt \, dV}_{inj}(z) = \left(\frac{n_{DM}}{2}\right) \, \left(\langle\sigma v \rangle n_{DM}\right)\,  \left( 2m_{DM} \right).
\label{Einj}
\end{equation}
\\
Assuming the dark matter particle to be non-relativistic and considering only the smooth cosmological dark matter halo, we find

\begin{equation}
n_{DM}= \rho_{DM} M_{DM}= \rho_c\, \Omega_{DM}\, M_{DM}= \rho_c\, \Omega_{DM \, o} \, (1+z)^3 \, M_{DM},
\label{eqnDM}
\end{equation}where $\Omega_{DM \, o}$ is the dark matter abundance today.\\

Substituting Eq.\eqref{eqnDM} into Eq.\eqref{Einj} we obtain

\begin{equation}
\frac{dE}{dt \, dV}_{inj}(z) = \rho_c^2 \, \Omega_{DM \, o}^2 \, (1+z)^6 \left( \frac{\langle\sigma v \rangle}{M_{DM}} \right).
\label{Einj}
\end{equation}

The injected energy is, however, not equal to the energy  deposited in the inter-galactic medium, because the response of the medium to energy injection does depend on the final states produced in the dark matter as well as when this took place. Hence, in order to account for this effect, we add an efficiency function which is redshift dependent, $f(z)$. In this way,
\begin{equation}
\frac{dE}{dt \, dV}_{dep} (z)= f(z)\frac{dE}{dt \, dV}_{inj}(z),
\end{equation}which means that the particle physics input of this effect is encompassed in $f(z) \langle\sigma v\rangle /M_{DM}$. However $f(z)$ is non-trivial to compute. Recently, the redshift dependence has been averaged over and the efficiency deposition has been separated from the energy spectrum \cite{Slatyer:2015jla}, as follows,

\begin{equation}
f_{eff}=\frac{1}{2 M_{DM}}\int_0^{M_{DM}}EdE\left(f_{eff}^{\gamma}(E)\frac{dN}{dE^{\gamma}}+2 f_{eff}^{e^+}(E)\frac{dN}{dE^{e^+}}\right),
\label{feeEq}
\end{equation}where the effect of dark matter annihilations on the CMB for any annihilation final states is determined by,

\begin{equation}
p_{ann}=f_{eff} \frac{\langle \sigma v \rangle}{m_{DM}}.
\label{annifactCMB}
\end{equation}

We highlight that $\langle \sigma v \rangle$ in Eq.\eqref{annifactCMB} is the thermally-averaged dark matter annihilation cross-section which can be very different from the dark matter annihilation cross-section today that enters into the  gamma-ray flux.\\

In Eq.\eqref{feeEq}, E is the energy of the electron-positron or photon, and $dN/dE^\gamma$ and $dN/dE^{e^+}$ are the photon and positron energy spectra produced per dark matter annihilation. Throughout we assume that the electrons and positrons have the same energy spectra. The energy and positron spectra produced per dark matter annihilation depends on the dark matter mass and annihilation final state. For instance, dark matter annihilations into VV which eventually decay into $e^+e^-$ produce an energy spectrum different from the direct annihilation into $e^+e^-$. In order to obtain CMB bounds for this $2\rightarrow 4$ annihilation process, we feed the energy spectrum obtained using Pythia \cite{Sjostrand:2014zea} into the code presented in \cite{Slatyer:2015jla,Slatyer:2015kla} who computed the new physics effects on the CMB.\\

Now we have reviewed the dark matter annihilation signatures in gamma-rays and for the CMB, we describe the dataset used in our analysis.

\begin{figure}[!h]
\centering
\includegraphics[width=1\columnwidth]{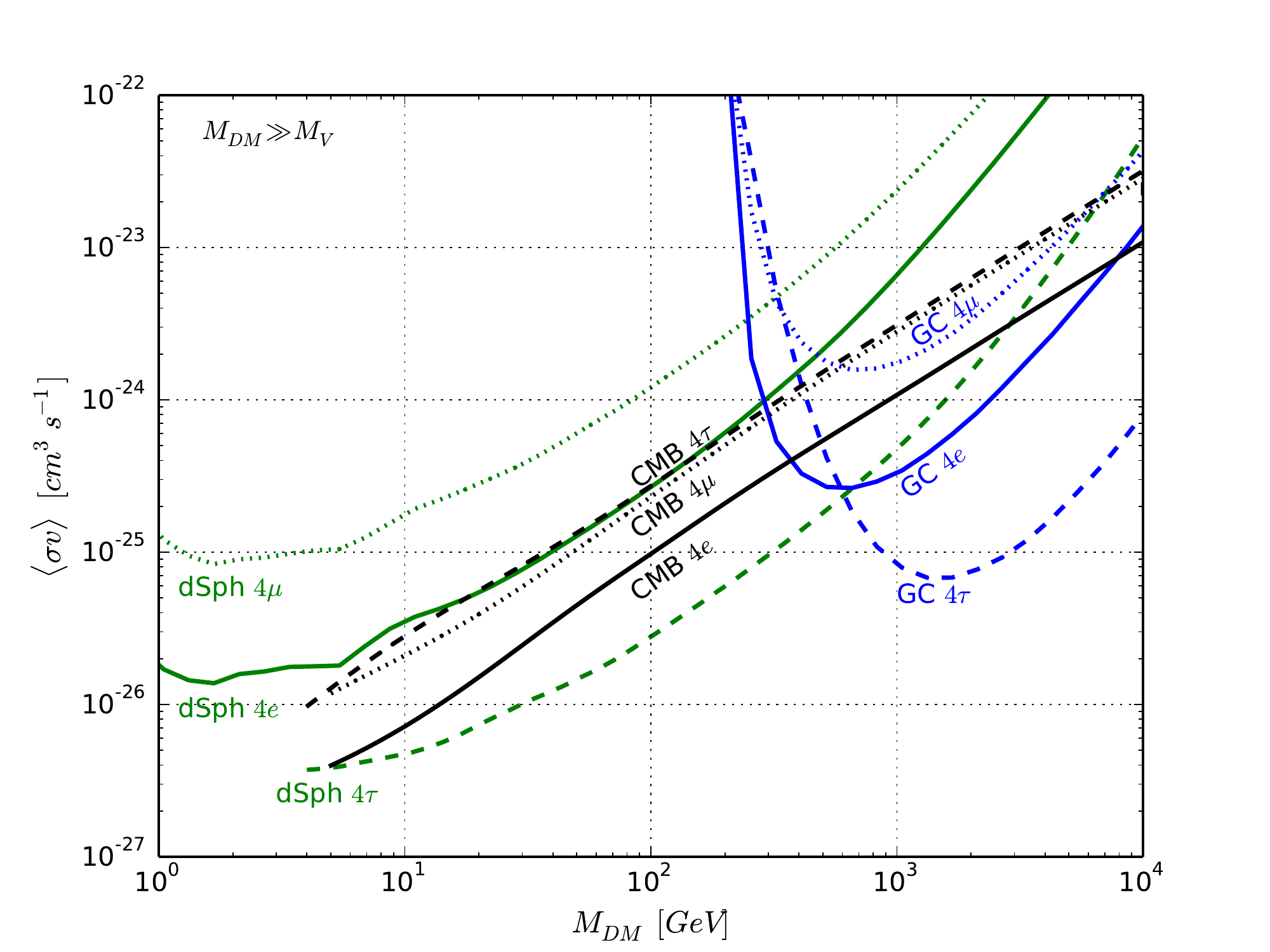}
\caption{95\% C.L. upper limits on the dark matter annihilation cross-section as a function of the dark matter mass for the process ${\rm DM \, DM} \rightarrow V V$, with $V$ decaying into $e\bar{e}$, $\mu\bar{\mu}$, $\tau\bar{\tau}$. We assumed $M_{DM} \gg M_V$, with the $V$ mass being just sufficiently heavier than the decays modes.}
\label{bound1}
\end{figure}

\section{Dataset}

\subsection{Fermi-LAT}

Fermi-LAT, the NASA gamma-ray telescope, has given rise to a new area of gamma-rays studies due to the precision achieved, far better than its predecessors, and for making the data publicly available to the community via user-friendly  tools \cite{Gehrels:1999ri}. In the context of dark matter, several important steps have been taken with the publication of several  studies of sensitivity to dark matter annihilations, for instance, in dSphs \cite{Abdo:2010ex,Ackermann:2011wa,Ackermann:2012nb,Ackermann:2013yva,Ackermann:2015zua,Drlica-Wagner:2015xua,Rico:2015nya,Ahnen:2016qkx}. \\

Since dSphs are dark matter-dominated objects and are located away from the large diffuse gamma-ray emission stemming from the galactic center, they are regarded as a ``holy grail'' as far as dark matter signals are concerned \cite{Blumenthal:1984bp,Peebles:1984zz,Dekek:1986gu,Kent:1987zz,Moore:1995pb,Ferrara:1999ry,Firmani:2000qe,vandenBosch:2000rza}. Be that as it may, there are still uncertainties on dark matter content in dSphs and these are subject of intense discussions \cite{Walker:2009zp,Pasetto:2010se,DelPopolo:2011cj,Charbonnier:2011ft,Jardel:2012am,Collins:2013eek,Laporte:2013fwa,Geringer-Sameth:2014yza,Adams:2014bda,Bonnivard:2015tta,Chiappo:2016xfs}. Therefore, in order to gain statistics and perform a more robust analysis, the datasets of dSph galaxies have been combined and treated using a maximum likelihood analysis \cite{Ackermann:2011wa}. After collecting seven years of gamma-ray data with energies between $500$~MeV-$500$~GeV belonging to the event class P8R2SOURCEV6 using the up-to-date software PASS-8, no significant excess has been observed leading to the exclusion of the thermal dark matter annihilation cross-section of $3\times 10^{-26}$ for dark matter masses below $100$~GeV, for annihilations solely into $b\bar{b}$ \cite{Ackermann:2015zua}.\\

The collaboration has publicly shared individual likelihood functions which one can use to perform their own independent analysis for a given dSph\footnote{http://www-glast.stanford.edu/pub\_data/1048/}. However, we emphasize that such likelihood functions are for individual dphs and since we are interested in repeating Fermi-LAT procedure in a more robust way using a stacked analysis of the sample of 15 dSphs used by Fermi-LAT, we have to build a combined likelihood function following the receipt provided in \cite{Ackermann:2015zua}. Moreover, we build a likelihood function to marginalize the statistical uncertainties on the J-factor with the J-factor likelihood function defined as,

\begin{eqnarray}
\mathcal{L}_J (J_i| J_{obs,i}, \sigma_i) &= & \frac{1}{\ln(10)J_{obs,i}\sqrt{2\pi} \sigma_i}\nonumber\\
& \times  & \exp \left\{ -\frac{(\log_{10}(J_i)-\log_{10}(J_{obs,i}))^2}{2\sigma_i^2} \right\}\nonumber\;,
\end{eqnarray}where $J_{obs,i}$ is the measured J-factor with statistical error $\sigma_i$ of a dSph $i$, while $J_i$ is the correct J-factor value.\\

Therefore, one can compute the joint-likelihood function by multiplying each individual likelihood, having in mind that the likelihood of an individual dSph $i$ is,

\begin{equation}
   \tilde{\mathcal{L}_i}(\mu,\theta_i = 
   \lbrace\alpha_i,J_i\rbrace |D_i)=\mathcal{L}_i(\mu,\theta_i|D_i)\mathcal{L}_J (J_i|J_{obs,i},\sigma_i)\;,
\end{equation}where $\mu$ accounts for particle physics input, dark matter mass and annihilation cross-section, $\theta_i$ for the set of nuisance parameters from the LAT study ($\alpha_i$) as well as the J-factors of the dSphs $J_i$, where $D_i$ is the gamma-ray data set used in the analysis. Thus, after performing a joint-likelihood function and accounting for the statistical errors on the J-factors with the likelihood function defined above we were able to do test statistics and obtain 95\% C.L. upper limits on the dark matter annihilation cross-section by finding a change in the log-likelihood of $2.71/2$ as described in \cite{Ackermann:2015zua}. We highlight were able to nicely reproduce Fermi-LAT limits as shown in \cite{Queiroz:2016zwd,Profumo:2016idl}. \\

\begin{figure}[!h]
\centering
\includegraphics[width=1\columnwidth]{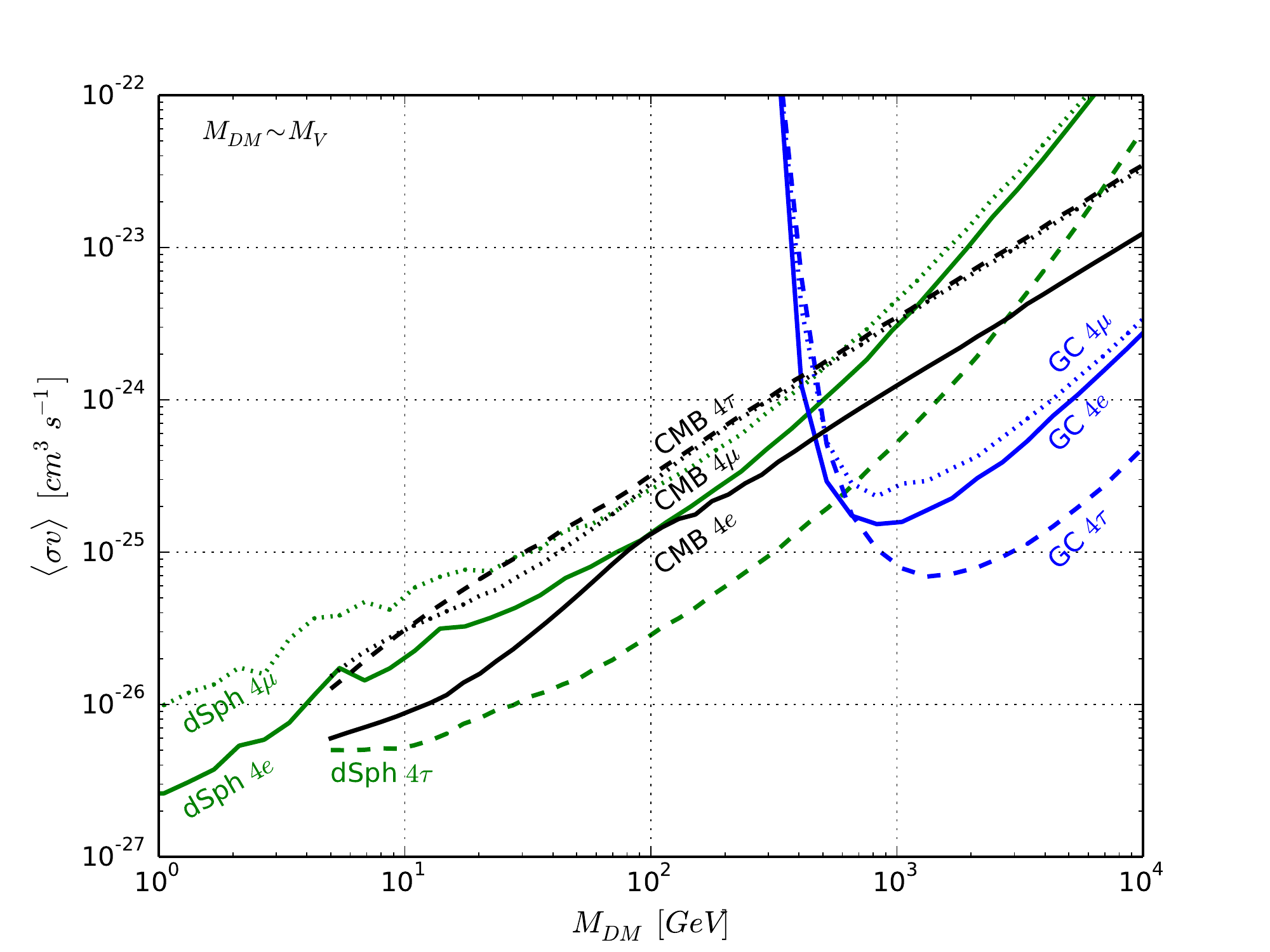}
\caption{95\% C.L. upper limits on the dark matter annihilation cross-section as a function of the dark matter mass for the process ${\rm DM \, DM} \rightarrow V V$, with $V$ decaying into $e\bar{e}$, $\mu \bar{\mu}$, $\tau \bar{\tau}$. We assumed $M_{DM} \sim M_V$.}
\label{bound2}
\end{figure}

In summary, we computed a joint-likelihood function for different energy spectra to take into account the fact we are now investigating secluded annihilations. The results are exhibit in green curves in Fig.\ref{bound1}-\ref{bound4} for two different mass regimes, $M_{DM} \gg M_V$ and $M_{DM} \sim M_V$.

\subsection{H.E.S.S.}
\label{sec:hess}

The H.E.S.S. telescope has expanded our knowledge concerning gamma-ray emitters for energies greater than $200$~GeV. In comparison to Fermi-LAT, H.E.S.S. lacks in exposure but overcomes this fact with its large effective area. As far as dark matter searches are concerned the most stringent limits on the dark matter annihilation cross-section come from observations of the Galactic Center. In particular, with 112h of data, upper limits were placed on the dark matter annihilation cross-section adopting an Einasto profile \cite{Abramowski:2011hc} still far from the thermal annihilation cross-section. A more recent study, with 10 years of observation that is equivalent to  254h of live-time exposure, H.E.S.S. has significantly improved their sensitivity, excluding the thermal annihilation cross-section for dark matter masses between $1-2$~TeV, annihilating solely into $\tau\bar{\tau}$ \cite{Abdallah:2016ygi}, again using a Einasto dark matter density profile. The substantial improvement on the bound is a result of more statistics and an improved analysis of the signal and background events \cite{Lefranc:2015vza}.\\

\begin{figure}[!h]
\centering
\includegraphics[width=1\columnwidth]{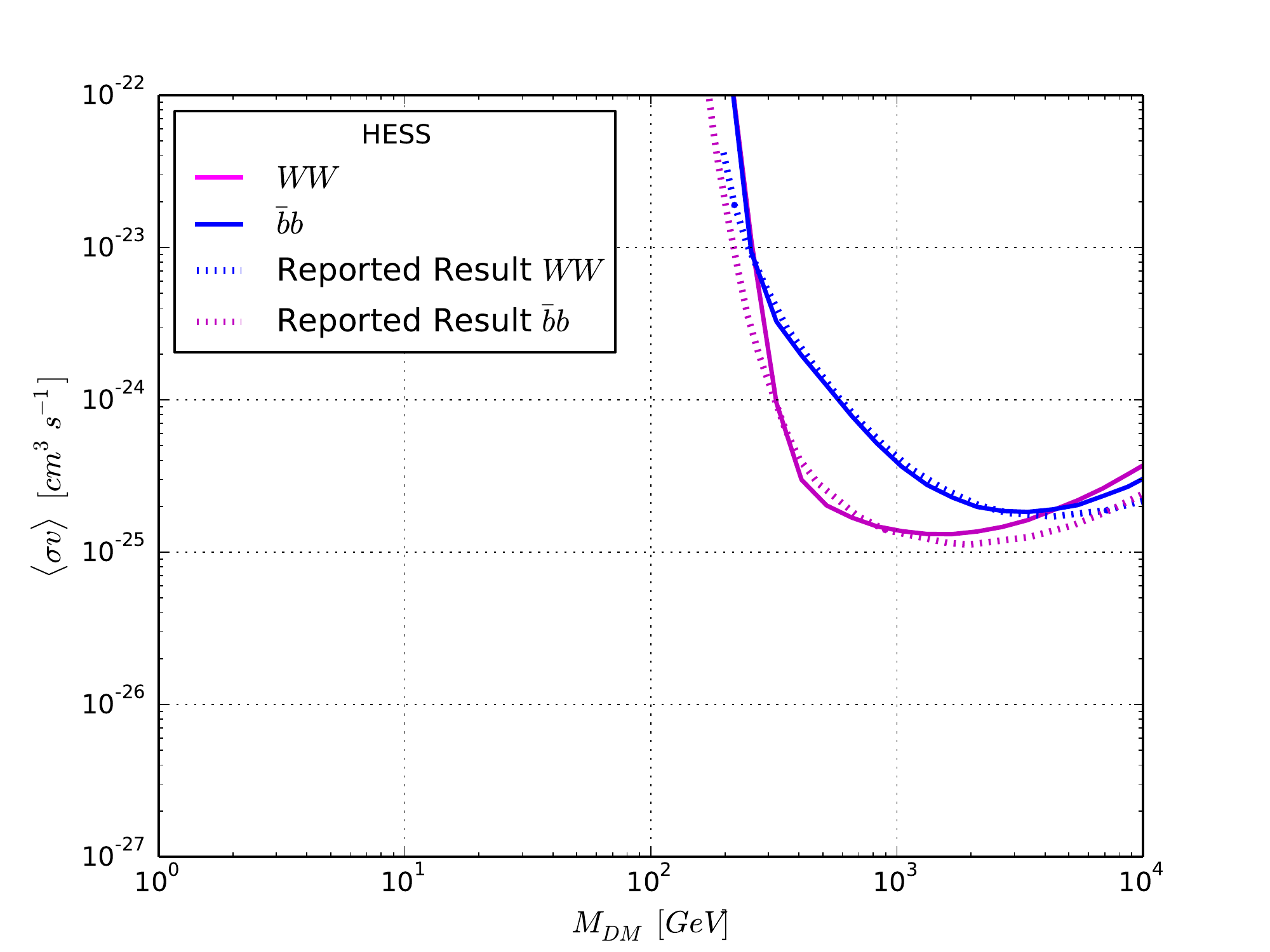}
\caption{Comparison between our results (continuous) and the results reported by H.E.S.S. Collaboration (dotted) for $\bar{b}b$ (blue) and $WW$ (magenta) channels \cite{Abdallah:2016ygi}. As one can see, we could reproduce their results in a more conservative way.}
\label{comparison}
\end{figure}

\begin{figure}[!h]
\centering
\includegraphics[width=1\columnwidth]{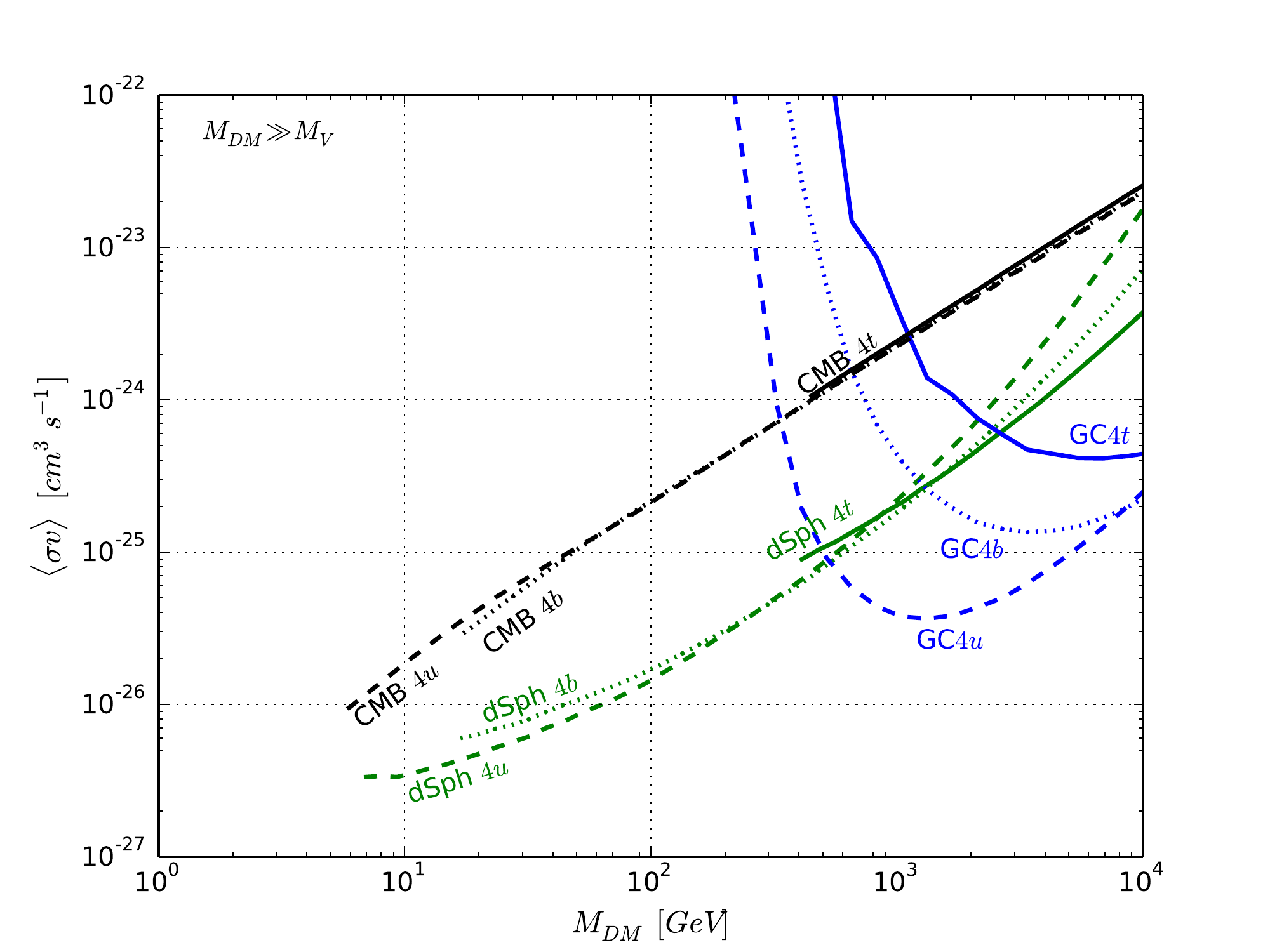}
\caption{95\% C.L. upper limits on the dark matter annihilation cross-section as a function of the dark matter mass for the process ${\rm DM \, DM} \rightarrow V V$, with $V$ decaying into $u\bar{u}$, $b \bar{b}$, $t \bar{t}$. We assumed $M_{DM} \gg M_V$, with the $V$ mass being just sufficiently heavier than the decays modes.}
\label{bound3}
\end{figure}

We duplicated the H.E.S.S. analysis and  were able to reproduce their results, as demonstrated in \cite{Profumo:2016idl}. We then fed the energy spectrum for secluded dark matter annihilations into the likelihood functions provided in \cite{Lefranc:2015vza} and changed the choice for the dark matter density profile to NFW, to put H.E.S.S. constraints on an equal footing with Fermi-LAT's. By performing a test-statistic as described in \cite{Lefranc:2015vza} we derive 95\% C.L. limits on the pair {\it dark matter annihilation cross section vs mass} for all channels here under consideration. The results are shown as blue curves in Fig.\ref{bound1}-\ref{bound4}  for various annihilation channels and two different mass regimes, $M_{DM} \gg M_V$ and $M_{DM} \sim M_V$.

\subsection{Planck}

The precision achieved by Planck has greatly impacted our knowledge about the Universe's constituents. The precise measurements of the CMB have resulted in a very precise estimate of the dark matter abundance and allowed us to make robust assessments on the impact of new physics effects on the CMB \cite{Ade:2013zuv,Planck:2013nga,Ade:2015xua}. \\

In particular, the precise measurements on the CMB temperature and polarization angular power spectra severely restrict potential new physics contributions that might change the ionization history \cite{Padmanabhan:2005es,Galli:2009zc,Galli:2011rz,Finkbeiner:2011dx,Madhavacheril:2013cna,Lopez-Honorez:2013lcm,Poulin:2015pna}. After integrating out all the cosmological factors, this affect can be described in terms of the annihilation factor $p_{ann}$ defined in Eq.\eqref{annifactCMB}. Currently, Planck places a 95\% C.L. limit on this parameter that reads, 

\begin{equation}
p_{ann} < 4.1 \times 10^{-28} cm^3 s^{-1} GeV^{-1} .
\label{CMBbound}
\end{equation}

\begin{figure}[!h]
\centering
\includegraphics[width=1\columnwidth]{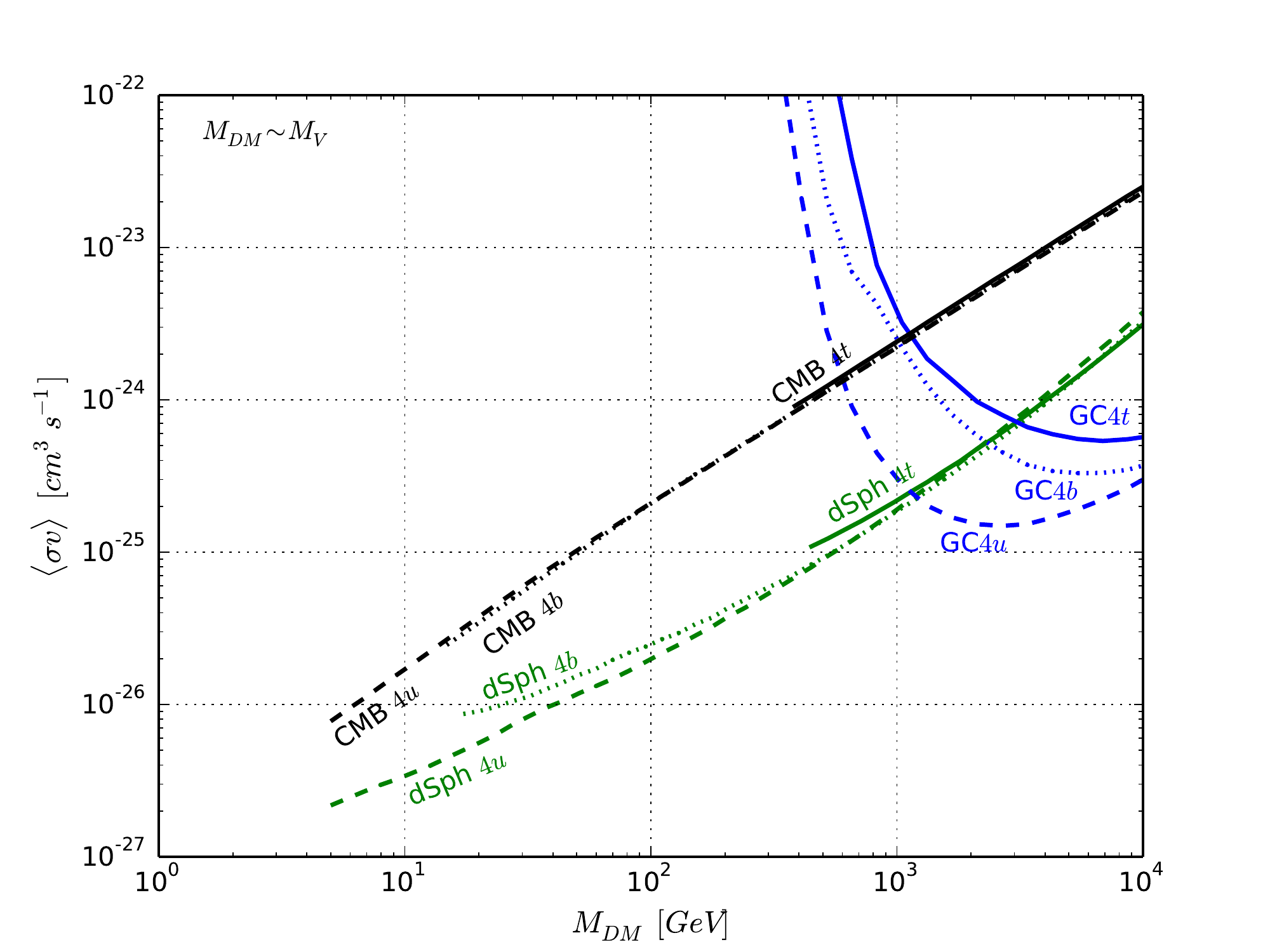}
\caption{95\% C.L. upper limits on the dark matter annihilation cross as a function of the dark matter mass for the process ${\rm DM DM} \rightarrow V V$, with $V$ decaying into $u\bar{u}$, $b \bar{b}$, $t \bar{t}$. We assumed $M_{DM} \sim M_V$.}
\label{bound4}
\end{figure}

Therefore, one can use this bound to constrain the ratio $\langle\sigma v \rangle/ M_{DM}$ for a given energy spectrum. Using the code in \cite{Slatyer:2015jla,Slatyer:2015kla}, we introduced the energy spectrum resulted from the various secluded dark matter scenarios we study, and compared to Eq.\eqref{CMBbound} to place 95\% C.L. on the dark matter annihilation cross-section. The limits are shown with black curves in Fig.\ref{bound1}-\ref{bound4} under two different mass regimes, $M_{DM} \gg M_V$ and $M_{DM} \sim M_V$.

\begin{table}
\begin{center}
{\color{Green} \bf \large Fermi-LAT PASS8}\\[0.5cm]
\begin{tabular}{|c|c|}
\hline
{\color{Green} \bf dSph 4b, $M_{DM}=100$~GeV } &  {\color{Green} \bf dSph 4b, $M_{DM}=1$~TeV}\\
$M_{DM}\gg M_V \rightarrow \sigma v \simeq 2 \times 10^{-26}\, cm^3/s$ & $M_{DM}\gg M_V \rightarrow \sigma v \simeq 2 \times 10^{-25}\, cm^3/s$  \\
$M_{DM}\sim M_V \rightarrow \sigma v \simeq 2.8 \times 10^{-26}\, cm^3/s$ & $M_{DM}\sim M_V \rightarrow \sigma v \simeq 2 \times 10^{-25}\, cm^3/s$  \\
\hline
{\color{Green} \bf dSph 4e, $M_{DM}=100$~GeV } &  {\color{Green} \bf dSph 4e, $M_{DM}=1$~TeV}\\
$M_{DM}\gg M_V \rightarrow \sigma v \simeq 3 \times 10^{-25}\, cm^3/s$ & $M_{DM}\gg M_V \rightarrow \sigma v \simeq 7 \times 10^{-24}\, cm^3/s$  \\
$M_{DM}\sim M_V \rightarrow \sigma v \simeq 1.5 \times 10^{-25}\, cm^3/s$ & $M_{DM}\sim M_V \rightarrow \sigma v \simeq 3.5 \times 10^{-24}\, cm^3/s$  \\
\hline
\end{tabular}\\
\vspace*{0.5cm}
{\color{blue} \bf \large H.E.S.S.}\\[0.5cm]
\begin{tabular}{|c|c|}
\hline
{\color{blue} \bf GC 4b, $M_{DM}=1$~TeV } &  {\color{blue} \bf GC 4b, $M_{DM}=10$~TeV}\\
$M_{DM}\gg M_V \rightarrow \sigma v \simeq  4.1 \times 10^{-25}\, cm^3/s$ & $M_{DM}\gg M_V \rightarrow \sigma v \simeq  2.0 \times 10^{-25}\, cm^3/s$  \\
$M_{DM}\sim M_V \rightarrow \sigma v \simeq  2.5 \times 10^{-24}\, cm^3/s$ & $M_{DM}\sim M_V \rightarrow \sigma v \simeq 3.6 \times 10^{-25}\, cm^3/s$  \\
\hline
{\color{blue} \bf GC $4 e$, $M_{DM}=1$~TeV } &  {\color{blue} \bf GC $4e$, $M_{DM}=10$~TeV}\\
$M_{DM}\gg M_V \rightarrow \sigma v \simeq  3.2 \times 10^{-25}\, cm^3/s$ & $M_{DM}\gg M_V \rightarrow \sigma v \simeq  1.3 \times 10^{-23}\, cm^3/s$  \\
$M_{DM}\sim M_V \rightarrow \sigma v \simeq 1.4 \times 10^{-25}\, cm^3/s$ & $M_{DM}\sim M_V \rightarrow \sigma v \simeq 2.4 \times 10^{-24}\, cm^3/s$  \\
\hline
{\color{blue} \bf GC $4\mu$, $M_{DM}=1$~TeV } &  {\color{blue} \bf GC $4\mu$, $M_{DM}=10$~TeV}\\
$M_{DM}\gg M_V \rightarrow \sigma v \simeq  1.7 \times 10^{-24}\, cm^3/s$ & $M_{DM}\gg M_V \rightarrow \sigma v \simeq  4.0 \times 10^{-23}\, cm^3/s$  \\
$M_{DM}\sim M_V \rightarrow \sigma v \simeq 2.5 \times 10^{-25}\, cm^3/s$ & $M_{DM}\sim M_V \rightarrow \sigma v \simeq 3.0 \times 10^{-24}\, cm^3/s$  \\
\hline
{\color{blue} \bf GC $4\tau$, $M_{DM}=1$~TeV } &  {\color{blue} \bf GC $4\tau$, $M_{DM}=10$~TeV}\\
$M_{DM}\gg M_V \rightarrow \sigma v \simeq  7.9 \times 10^{-26}\, cm^3/s$ & $M_{DM}\gg M_V \rightarrow \sigma v \simeq  7.4 \times 10^{-25}\, cm^3/s$  \\
$M_{DM}\sim M_V \rightarrow \sigma v \simeq 7.6 \times 10^{-26}\, cm^3/s$ & $M_{DM}\sim M_V \rightarrow \sigma v \simeq 4.4 \times 10^{-25}\, cm^3/s$  \\
\hline
\end{tabular}\\
\vspace*{0.5cm}
{\color{black} \bf \large Planck}\\[0.5cm]
\begin{tabular}{|c|c|}
\hline
{\color{black} \bf CMB 4b, $M_{DM}=100$~GeV } &  {\color{black} \bf CMB 4b, $M_{DM}=1$~TeV}\\
$M_{DM}\gg M_V \rightarrow \langle \sigma v\rangle \simeq 2.5  \times 10^{-25}\, cm^3/s$ & $M_{DM}\gg M_V \rightarrow \langle \sigma v \rangle\simeq 3 \times 10^{-24}\, cm^3/s$  \\
$M_{DM}\sim M_V \rightarrow \langle \sigma v \rangle \simeq 2.5 \times 10^{-25}\, cm^3/s$ & $M_{DM}\sim M_V \rightarrow \langle \sigma v\rangle \simeq 2.5 \times 10^{-24}\, cm^3/s$  \\
\hline
{\color{black} \bf CMB 4e, $M_{DM}=100$~GeV } &  {\color{black} \bf CMB 4e, $M_{DM}=1$~TeV}\\
$M_{DM}\gg M_V \rightarrow \langle\sigma v\rangle \simeq  10^{-25}\, cm^3/s$ & $M_{DM}\gg M_V \rightarrow \langle\sigma v\rangle \simeq  10^{-24}\, cm^3/s$  \\
$M_{DM}\sim M_V \rightarrow \langle\sigma v\rangle \simeq 1.5 \times 10^{-25}\, cm^3/s$ & $M_{DM}\sim M_V \rightarrow \langle\sigma v\rangle \simeq 1.5 \times 10^{-24}\, cm^3/s$  \\
\hline
\end{tabular}
\end{center}
\caption{95\% C.L. limits on the dark matter annihilation cross-section for several secluded dark matter scenarios. For easy comparison, we show the bounds for $M_{DM} \gg M_V$ and $M_{DM} \sim M_V$. All annihilations were individually analyzed, i.e. we assumed that the dark matter particle annihilates solely into each final state. Notice that Fermi-LAT and H.E.S.S. are sensitive to the dark matter annihilation cross-section today, whereas CMB is sensitive to the thermally averaged cross-section.}
\label{table2}
\end{table}

\section{Results}
\label{sec:results}

In Fig.\ref{bound1} we exhibit the 95\% C.L. limits on the dark matter annihilation cross-section for the case that $M_{DM} \gg M_V$, with the mediator being just sufficiently heavier than its decay channels. We remind the Reader that the CMB bounds are placed on the thermally-averaged dark matter annihilation cross-section, whereas Fermi-LAT and H.E.S.S. bounds are on the dark matter annihilation cross-section today. In the figures, we keep the average dark matter annihilation on the y-axis, but the reader should interpret the plots with caution.\\

The bounds from CMB are shown in black lines for the $4e$ (solid), $4 \mu$(dotted) and $4\tau$ (dashed) channels. A similar pattern is used for Fermi-LAT (green) and H.E.S.S. (blue) bounds.\\

As expected, the  CMB is a powerful probe of dark matter annihilations into electron and muons. For annihilations into VV, with each V decaying into $\bar{\tau}\tau$ Fermi-LAT takes the lead since $\tau$ decays hadronically and efficiently produces gamma-rays. For dark matter masses above $400$~GeV, the complementarity among these three telescopes is very encouraging. For annihilations into $4e$ and $4\mu$, the CMB is still more restrictive, but the $4\tau$ Fermi-LAT bound is the most stringent for $M_{DM} < 800$~GeV, with H.E.S.S. taking the lead for larger masses.\\

Similarly to what has been typically assumed in the literature, we adopted $M_{DM} \gg M_V$. We now assume that the dark matter and the mediator are nearly 
mass-degenerate and show the new bounds on the annihilation cross-section in Fig.\ref{bound2}. Our choice for this assumption of a mass-degenerate regime is motivated by its presence in several dark sector models. On a theoretical note, if the dark matter particle has both vector and axial vector couplings to the mediator then we anticipate that the dark matter particle cannot be much heavier than the mediator, so as not to spoil perturbativity unitarity \cite{Shu:2007wg,Kahlhoefer:2015bea}. Thus  it is reasonable to assume that the dark matter and the mediator have comparable masses. \\

H.E.S.S. sensitivity to annihilations into $4\tau$ is not much affected, but those into $4e$ and $4\mu$ are visibly changed. For instance, with $M_{DM}=1$~TeV, for $e\bar{e}$ final state, H.E.S.S. could exclude $\sigma v \simeq 3.2 \times 10^{-25} cm^3/s$ for $M_{DM} \gg M_{V}$, but with  $M_{DM} \sim M_{V}$ now excludes $\sigma v \simeq 1.4 \times 10^{-25} cm^3/s$. Moreover, still keeping $M_{DM}=1$~TeV, H.E.S.S. upper bound on dark matter annihilation cross-sections, for $\mu\bar{\mu}$ final state, goes from  $\sigma v \simeq 1.7 \times 10^{-24} cm^3/s$ for $M_{DM} \gg M_V$, to $\sigma v \simeq 2.5 \times 10^{-25} cm^3/s$ for $M_{DM} \sim M_{V}$, clearly showing that the mediator mass is important in the derivation of robust limits on the dark matter annihilation cross-section. A similar reasoning can be applied to the other experiments. On Table \ref{table2} we show several benchmark models for comparison.\\

In Fig.\ref{bound3} we exhibit the 95\% C.L. limits on the dark matter annihilation cross-section into hadrons for the case that $M_{DM} \gg M_V$, with the mediator being just sufficiently heavier than its decay final states. The CMB bounds are shown in black lines for the annihilations into $4 t$ (solid), $4 b$ (dotted) and $4 u$ (dashed) channels. A similar pattern is used for Fermi-LAT (green) and H.E.S.S. (blue) bounds.\\

For $M_{DM}=1$~TeV and annihilations into $4b$ with $M_{DM} \gg M_V$, Fermi-LAT excludes $\sigma v \simeq 2 \times 10^{-25} cm^3/s$, Planck rules out $\langle\sigma v \rangle \simeq 3 \times 10^{-24} cm^3/s$, and lastly H.E.S.S. excludes $\sigma v \simeq 4.1 \times 10^{-25} cm^3/s$. If we ramp up the mediator mass to $M_V \sim M_{DM}$, the H.E.S.S. sensitivity significantly changes, whereas Planck and Fermi-LAT sensitivities are mildly changed. In particular, the H.E.S.S. upper bound on the dark matter annihilation cross-section worsens now ruling out $\sigma v \simeq 2.5 \times 10^{-24} cm^3/s$. A multitude of other setups are explicitly presented in Table \ref{table2} to facilitate the comparisons and easily assess the impact of the mediator mass in the derivation of the bounds.\\

\begin{figure}[!h]
\centering
\includegraphics[width=0.49\columnwidth]{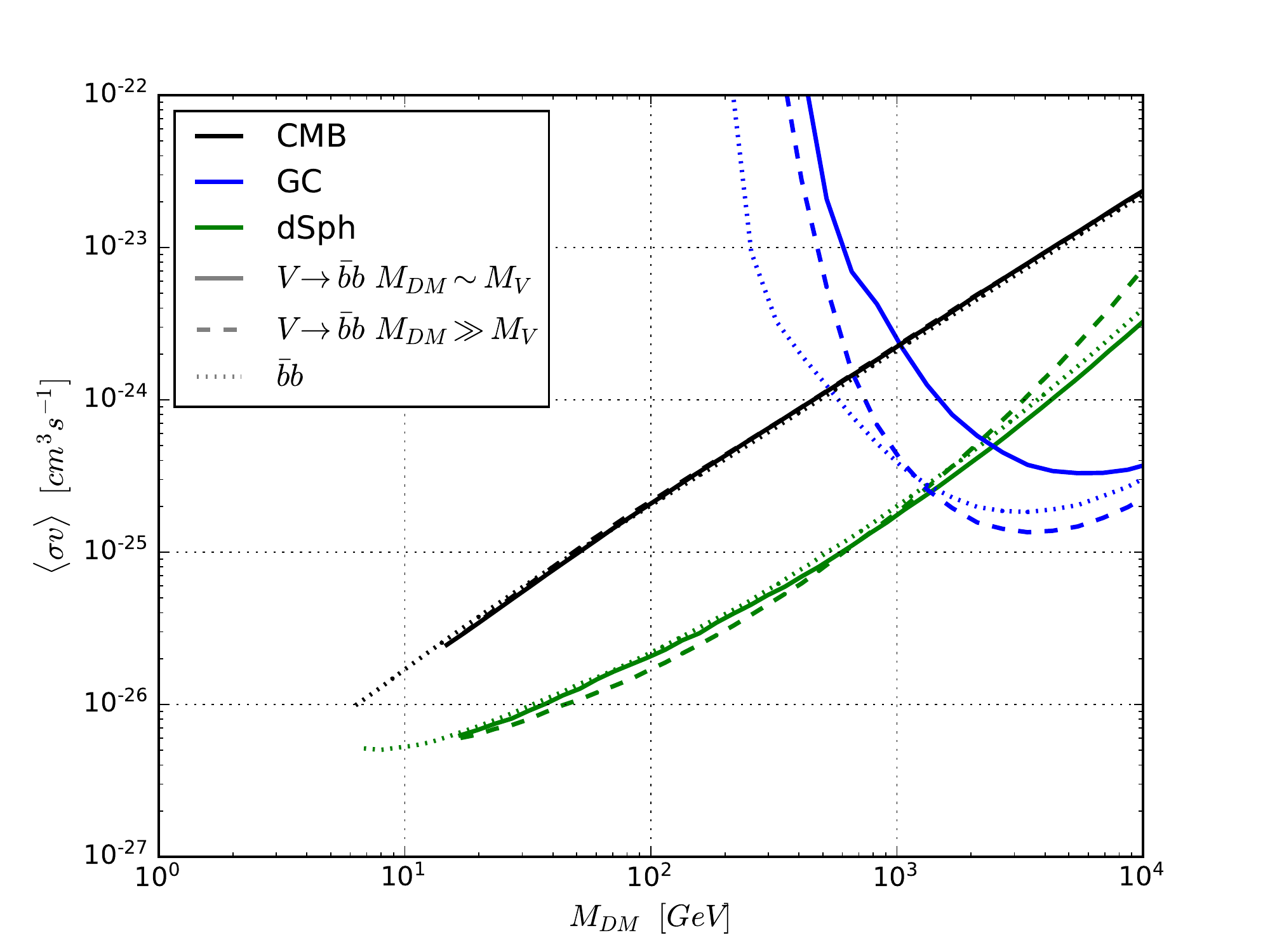}
\includegraphics[width=0.49\columnwidth]{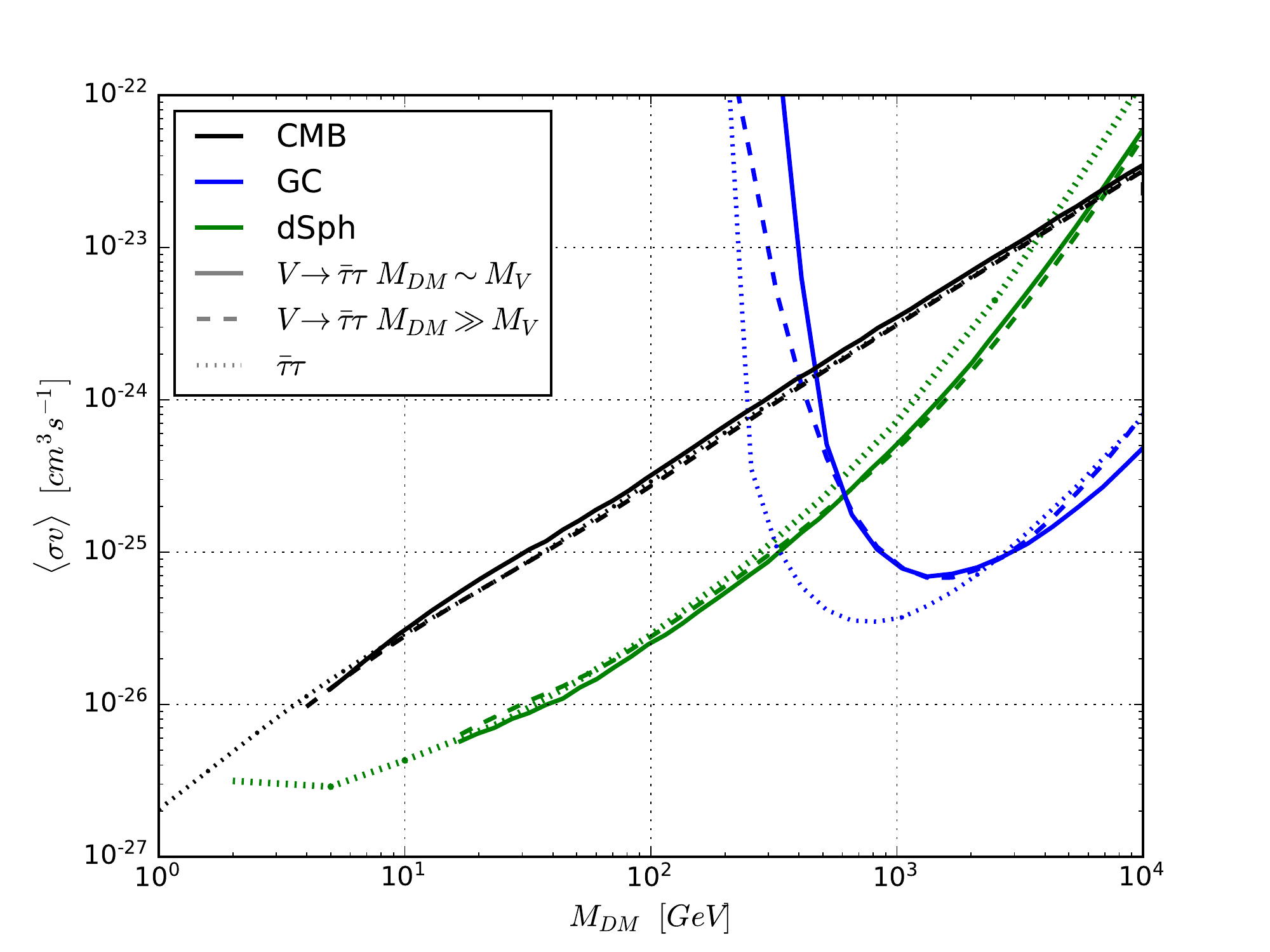}
\caption{95\% C.L. upper limits on the dark matter annihilation cross section for $M_{DM} \gg M_V$ and $M_{DM} \sim M_V$ to easy comparison based on the observation from the CMB (black), GC (blue) and dSph (green). {\it Left-panel:} annihilation into $4b$; {\it Right-panel:} annihilation into $4\tau$. We also overlay the canonical dark matter annihilation into $2b$ and $2\tau$ respectively. }
\label{boundcomparison}
\end{figure}

In summary, one can see the mediator mass is more relevant for  heavy dark matter, with masses much larger than $100$~GeV. In some cases, the difference in the upper limit on the dark matter annihilation cross-section reaches a factor of seven, as in the $M_{DM}=1$~TeV case for annihilations into $4\mu$, for the H.E.S.S. telescope. To illustrate even further the importance of the mediator mass we paired the results for $M_{DM} \gg M_V$ and $M_{DM} \sim M_V$ for the $4b$ and $4\tau$ channels in Fig.\ref{boundcomparison}. As noted previously, the upper limits change significantly specially for H.E.S.S. observation of the GC.\\

To summarize, our work derives  new and updated limits on the dark matter annihilation cross-section for the  secluded dark matter sector, and shows that these bounds rely on the mass of the mediator.

\section*{Discussion}

CMB and gamma-ray constraints do not probe the same annihilation rates, as they probe annihilation rates at different times in the history of the universe. The annihilation cross-section today drives gamma-ray signals, whereas the thermal averaged annihilation cross-section at CMB decoupling drives CMB constraints. Some dark matter models feature velocity-dependent annihilation rates, and in these models the assessment of which experiment is the most restrictive is model-dependent \cite{Robertson:2009bh,Campbell:2010xc,Hisano:2011dc,Arina:2014yna,Zhao:2016xie,Goncalves:2016iyg,Gonzalez-Morales:2017jkx}. Here, we have highlighted  that  bounds on the dark matter annihilation cross-section do not depend on  specific particle physics models and in this respect they can be regarded as model-independent.  

\section{TeV Gamma-ray Excess Observed by H.E.S.S.}

Gamma ray observations performed by the ground-based H.E.S.S. telescope within the $15$~pc from the Galactic Center revealed a very high energy point-like source known as J1745-290 which features a significant deviation from a pure power-law (PL) emission in the few TeV energy region \cite{Aharonian:2009zk,Abramowski:2016mir}.\\

\begin{figure}[!h]
\centering
\includegraphics[width=0.49\columnwidth]{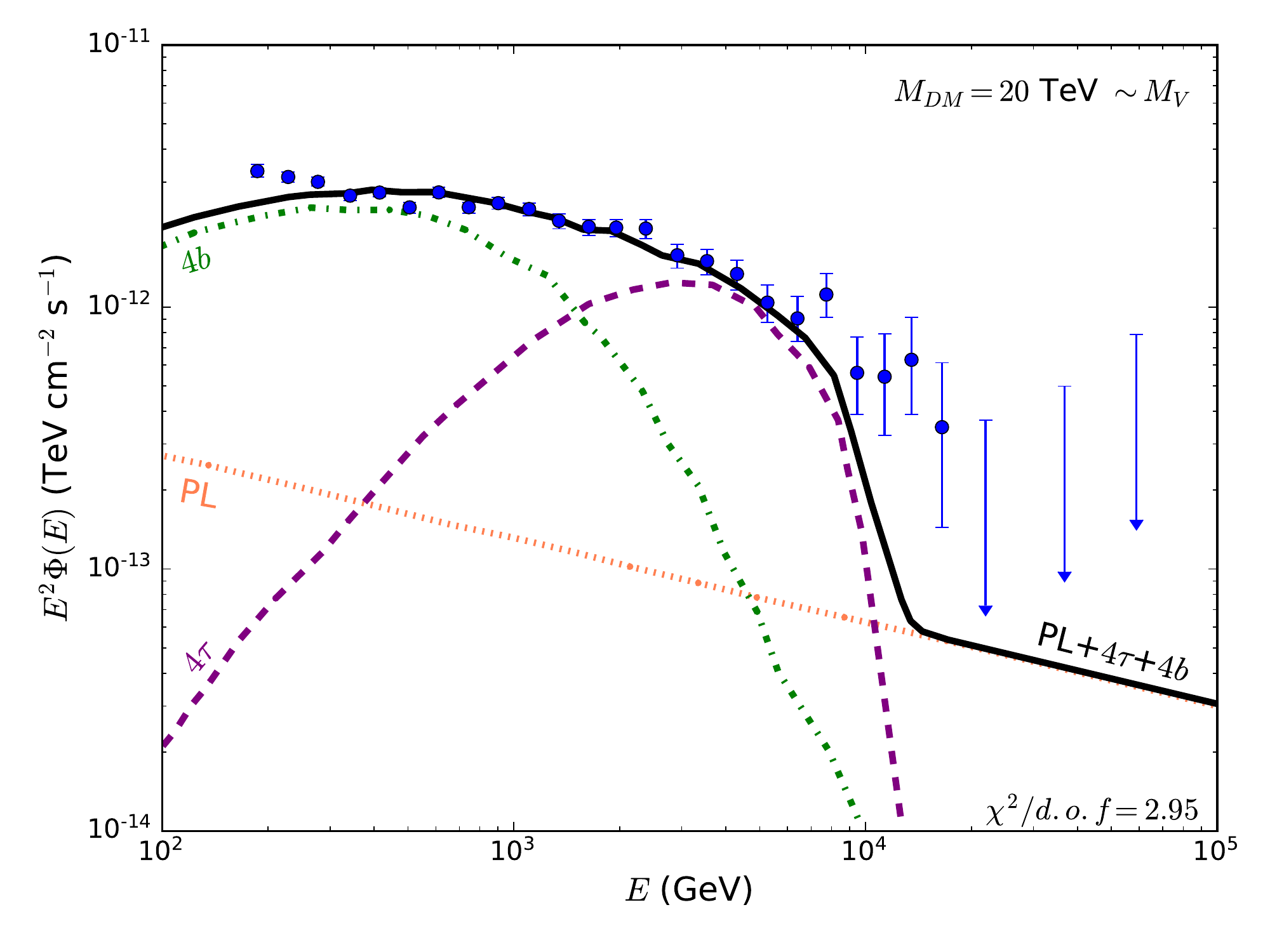}
\includegraphics[width=0.49\columnwidth]{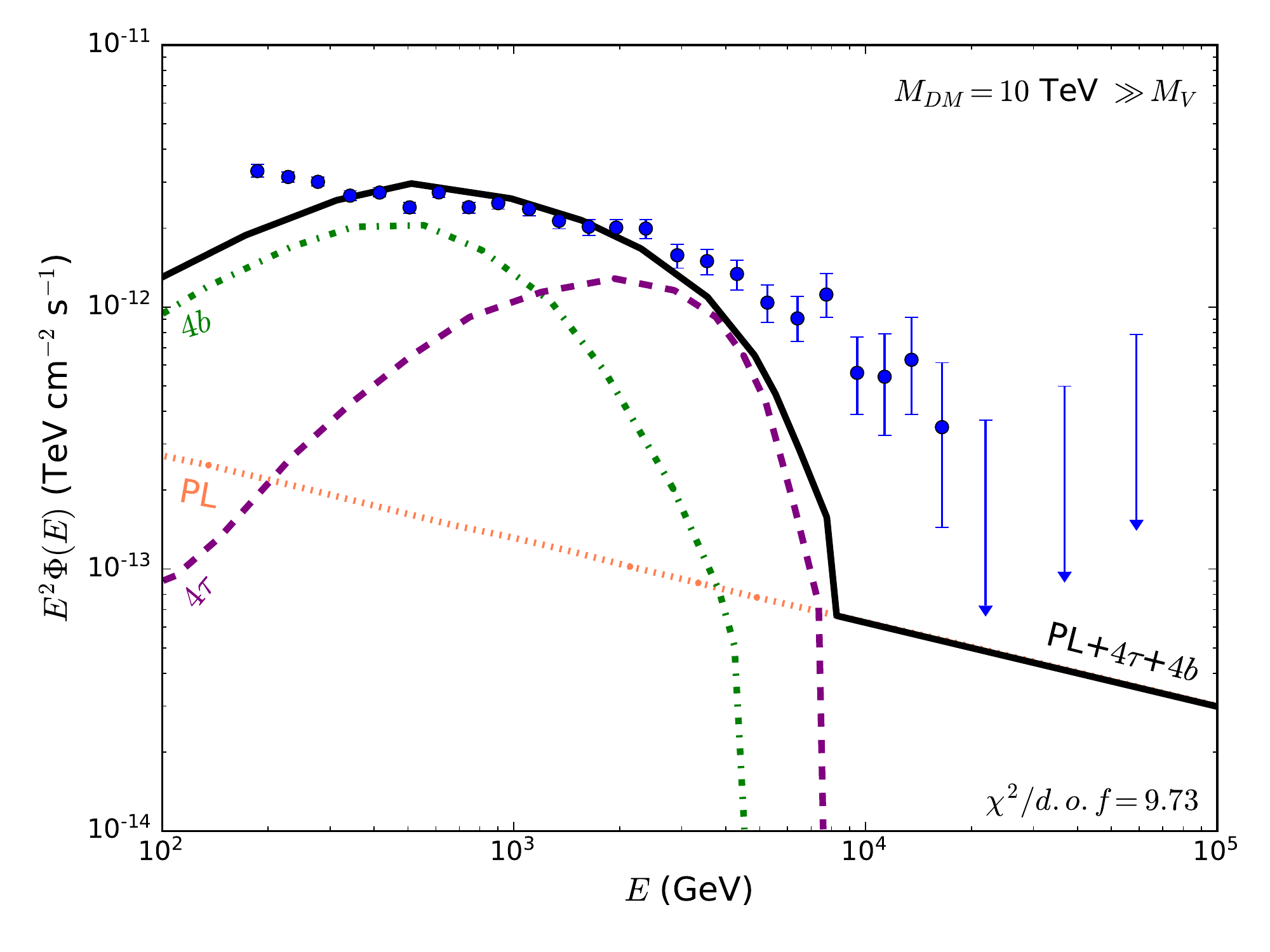}
\caption{H.E.S.S. TeV gamma-ray observation within $15$~pc of the Galactic Center. In blue are the data points. The dotted coral line is the background emission assumed to be a simple PL. In black in the total gamma-ray emission obtained by summing the dark matter component with the background. In green (purple), the gamma-ray emission produced by the dark matter particle annihilating into $4b$ ($4\tau$). {\it Left-panel}: 20 TeV dark matter particle, with $\sigma v = 3.87\times 10^{-26} cm^3 s^{-1}$, annihilating with 60\% branching ratio into $4b$ and 40\% into $4\tau$ for $M_{DM} \sim M_V$. {\it Right-panel}:  10 TeV DM particle, with $\sigma v = 1.64\times 10^{-26} cm^3 s^{-1}$, annihilating 50\% into $4b$ and 50\% into $4\tau$  for $M_{DM} \gg M_V$. One can quantitatively see the impact of the mediator mass.}
\label{figHESSexcess}
\end{figure}

We followed the procedure described in \cite{Abramowski:2016mir} and assumed that the annular region within $0.15^o$ and $0.45^o$ that has a solid angle of $1.4 \times 10^{-4}$~sr is well-fitted by the PL spectrum $dN/dE= A_b (E/1{\rm TeV})^{\Gamma_b}$, with $A_b= (1.92 \pm 0.08_{stat} \pm 0.28_{sys})\times 10^{-12} {\rm  cm^{-2} s^{-1} TeV^{-1} }$ and $\Gamma_b = 2.32 \pm 0.05_{stat} \pm 0.11_{syst}$. The result is shown with a dotted coral line in Fig. \ref{figHESSexcess}.\\

Moreover, we generated the energy spectrum of secluded dark matter annihilations using Pythia 8 \cite{Sjostrand:2014zea} and derived the gamma-ray fluxes for dark matter annihilations into $4\tau$ e $4b$ for $M_{DM} \sim M_V$ (left-panel) and $M_{DM} \gg M_V$ (right-panel), as displayed in in Fig.\ref{figHESSexcess} with green and purple curves, respectively. \\

After summing the background contribution to the power-law emission above, we obtain the black curve which is the gamma-ray flux produced by the dark matter plus background. In the left panel, where $M_{DM} \sim M_V$, the minimum $\chi^2$ procedure yields $\chi^2/d.o.f=2.95$. It favors $\sigma v =  3.87\times 10^{-26} cm^3 s^{-1}$ and $M_{DM}=20$~TeV with an annihilation branching ratio of 60\% into $4b$ and 40\% into $4\tau$. We also investigated the possibility to  improve the $\chi^2$ for different branching ratios but the aforementioned scenario provided the best-fit.\\

In the right panel, where $M_{DM} \gg M_V$, the minimum $\chi^2$ procedure yields $\chi^2/d.o.f=9.73$. It favors $\sigma v =  1.64\times 10^{-26} cm^3 s^{-1}$ and $M_{DM}=10$~TeV with the branching ratio of 50\% into $4b$ and 50\% into $4\tau$.\\

It is important to highlight that the dark matter annihilation cross-section serves as an overall rescaling of the energy spectrum, whereas the dark matter mass changes the energy spectrum behavior at lower and larger energies. Furthermore, dark matter annihilations in the mass degenerate case lead to a harder gamma-ray spectrum at lower energies proving a better fit to the data where the error is relatively smaller. The trade-off between dark matter mass, annihilation cross-section, annihilation final states led to the best-fit parameters found above.\\

In summary, the TeV gamma-ray excess observed the H.E.S.S. telescope is also well-fitted by secluded dark matter annihilations, and mediator mass has an important impact on what dark matter mass provides a best-fit to the data, ratifying our conclusions. \\

A similar study but in the context of $2\rightarrow 2$ dark matter annihilations was carried out in \cite{Belikov:2012ty,Belikov:2013laa}. It is important to emphasize though that one could also nicely explain the H.E.S.S. observation by assuming that the background emission departs from the pure power-law spectrum, and is actually described by a power-law function with an exponential cut-off \cite{Belikov:2016fwv}.

\section{Conclusions}
\label{sec:con}

Secluded dark sectors are often investigated in a model-dependent fashion. Motivated by their relevance, we used seven years of Fermi-LAT observations of dwarf spheroidal galaxies, 10 years of H.E.S.S. data from the  Galactic center, and the most recent Planck measurements on the Cosmic Microwave Background to constrain secluded dark sectors. We showed here that these datasets are highly complementary to each other, especially when the dark matter mass is above $\sim 400$~GeV. We derived 95\% C.L. limits on the dark matter annihilation cross-section for several scenarios. We sifted annihilations into $4e,4\mu,4\tau, 4u, 4b, 4t$ and also assessed the impact of the mediator mass on our bounds. As far as indirect searches for dark matter are concerned, these mediators are often assumed to be very light and just sufficiently heavier than the decay modes' particle masses. We showed that their masses can lead to sizeable changes on the overall upper limits on the dark matter annihilation 
cross-section. Our bounds supersede previous constraints using Fermi-LAT data, and they constitute the first limits on secluded dark sectors using the H.E.S.S. telescope. We also showed that secluded dark matter can also fit the TeV gamma-ray observation observed by H.E.S.S., with the mediator mass driving the location of the best-fit points. \\

In conclusion, our results clearly show that any robust assessment of secluded dark sectors in the context of indirect detection relies not only the dark matter annihilation cross-section and annihilation final states, but also on the mediator mass.

\section*{Acknowledgments}

We are grateful to the Fermi-LAT team for the publicly available likelihood functions used in this study. The authors thank Marco Cirelli and Tathagata Gosh for correspondence. We are also grateful to Giorgio Arcadi, Carlos Yaguna, Stefan Vogl, Yann Mambrini, Miguel Campos, Aion Viana, and Torsten Bringmann for discussions. We thank Filippo Sala for reminding us of the updated limits from HESS collaboration that were missing in a earlier version. FSQ acknowledges support from MEC and ICTP-SAIFR FAPESP grant 2016/01343-7. CS is supported by CAPES/PDSE Process 88881.134759/2016-01.

\bibliographystyle{JHEPfixed}
\bibliography{darkmatter}

\providecommand{\href}[2]{#2}\begingroup\raggedright\begin{thebibliography}{100}

\bibitem{Bertone:2004pz}
G.~Bertone, D.~Hooper, and J.~Silk, {\it {Particle dark matter: Evidence,
  candidates and constraints}},  {\em Phys. Rept.} {\bf 405} (2005) 279--390,
  [\href{http://xxx.lanl.gov/abs/hep-ph/0404175}{{\tt hep-ph/0404175}}].

\bibitem{Strigari:2013iaa}
L.~E. Strigari, {\it {Galactic Searches for Dark Matter}},  {\em Phys. Rept.}
  {\bf 531} (2013) 1--88, [\href{http://xxx.lanl.gov/abs/1211.7090}{{\tt
  1211.7090}}].

\bibitem{Profumo:2013yn}
S.~Profumo, {\it {Astrophysical Probes of Dark Matter}},  in {\em {Proceedings,
  Theoretical Advanced Study Institute in Elementary Particle Physics:
  Searching for New Physics at Small and Large Scales (TASI 2012): Boulder,
  Colorado, June 4-29, 2012}}, pp.~143--189, 2013.
\newblock \href{http://xxx.lanl.gov/abs/1301.0952}{{\tt 1301.0952}}.

\bibitem{Bertone:2016nfn}
G.~Bertone and D.~Hooper, {\it {A History of Dark Matter}},  {\em Submitted to:
  Rev. Mod. Phys.} (2016) [\href{http://xxx.lanl.gov/abs/1605.04909}{{\tt
  1605.04909}}].

\bibitem{Queiroz:2016sxf}
F.~S. Queiroz, W.~Rodejohann, and C.~E. Yaguna, {\it {Is the dark matter
  particle its own antiparticle?}},
  \href{http://xxx.lanl.gov/abs/1610.06581}{{\tt 1610.06581}}.

\bibitem{Arcadi:2017kky}
G.~Arcadi, M.~Dutra, P.~Ghosh, M.~Lindner, Y.~Mambrini, M.~Pierre, S.~Profumo,
  and F.~S. Queiroz, {\it {The Waning of the WIMP? A Review of Models,
  Searches, and Constraints}},  \href{http://xxx.lanl.gov/abs/1703.07364}{{\tt
  1703.07364}}.

\bibitem{Bringmann:2012ez}
T.~Bringmann and C.~Weniger, {\it {Gamma Ray Signals from Dark Matter:
  Concepts, Status and Prospects}},  {\em Phys. Dark Univ.} {\bf 1} (2012)
  194--217, [\href{http://xxx.lanl.gov/abs/1208.5481}{{\tt 1208.5481}}].

\bibitem{Bringmann:2011ye}
T.~Bringmann, F.~Calore, G.~Vertongen, and C.~Weniger, {\it {On the Relevance
  of Sharp Gamma-Ray Features for Indirect Dark Matter Searches}},  {\em Phys.
  Rev.} {\bf D84} (2011) 103525, [\href{http://xxx.lanl.gov/abs/1106.1874}{{\tt
  1106.1874}}].

\bibitem{Hooper:2012sr}
D.~Hooper, C.~Kelso, and F.~S. Queiroz, {\it {Stringent and Robust Constraints
  on the Dark Matter Annihilation Cross Section From the Region of the Galactic
  Center}},  {\em Astropart. Phys.} {\bf 46} (2013) 55--70,
  [\href{http://xxx.lanl.gov/abs/1209.3015}{{\tt 1209.3015}}].

\bibitem{Ibarra:2013zia}
A.~Ibarra, A.~S. Lamperstorfer, and J.~Silk, {\it {Dark matter annihilations
  and decays after the AMS-02 positron measurements}},  {\em Phys. Rev.} {\bf
  D89} (2014), no.~6 063539, [\href{http://xxx.lanl.gov/abs/1309.2570}{{\tt
  1309.2570}}].

\bibitem{Bergstrom:2013jra}
L.~Bergstrom, T.~Bringmann, I.~Cholis, D.~Hooper, and C.~Weniger, {\it {New
  limits on dark matter annihilation from AMS cosmic ray positron data}},  {\em
  Phys. Rev. Lett.} {\bf 111} (2013) 171101,
  [\href{http://xxx.lanl.gov/abs/1306.3983}{{\tt 1306.3983}}].

\bibitem{Lin:2014vja}
S.-J. Lin, Q.~Yuan, and X.-J. Bi, {\it {Quantitative study of the AMS-02
  electron/positron spectra: Implications for pulsars and dark matter
  properties}},  {\em Phys. Rev.} {\bf D91} (2015), no.~6 063508,
  [\href{http://xxx.lanl.gov/abs/1409.6248}{{\tt 1409.6248}}].

\bibitem{Bringmann:2014lpa}
T.~Bringmann, M.~Vollmann, and C.~Weniger, {\it {Updated cosmic-ray and radio
  constraints on light dark matter: Implications for the GeV gamma-ray excess
  at the Galactic center}},  {\em Phys. Rev.} {\bf D90} (2014), no.~12 123001,
  [\href{http://xxx.lanl.gov/abs/1406.6027}{{\tt 1406.6027}}].

\bibitem{Gonzalez-Morales:2014eaa}
A.~X. Gonzalez-Morales, S.~Profumo, and F.~S. Queiroz, {\it {Effect of Black
  Holes in Local Dwarf Spheroidal Galaxies on Gamma-Ray Constraints on Dark
  Matter Annihilation}},  {\em Phys. Rev.} {\bf D90} (2014), no.~10 103508,
  [\href{http://xxx.lanl.gov/abs/1406.2424}{{\tt 1406.2424}}].

\bibitem{DiMauro:2015jxa}
M.~Di~Mauro, F.~Donato, N.~Fornengo, and A.~Vittino, {\it {Dark matter vs.
  astrophysics in the interpretation of AMS-02 electron and positron data}},
  {\em JCAP} {\bf 1605} (2016), no.~05 031,
  [\href{http://xxx.lanl.gov/abs/1507.07001}{{\tt 1507.07001}}].

\bibitem{Buckley:2015doa}
M.~R. Buckley, E.~Charles, J.~M. Gaskins, A.~M. Brooks, A.~Drlica-Wagner,
  P.~Martin, and G.~Zhao, {\it {Search for Gamma-ray Emission from Dark Matter
  Annihilation in the Large Magellanic Cloud with the Fermi Large Area
  Telescope}},  {\em Phys. Rev.} {\bf D91} (2015), no.~10 102001,
  [\href{http://xxx.lanl.gov/abs/1502.01020}{{\tt 1502.01020}}].

\bibitem{Giesen:2015ufa}
G.~Giesen, M.~Boudaud, Y.~Génolini, V.~Poulin, M.~Cirelli, P.~Salati, and
  P.~D. Serpico, {\it {AMS-02 antiprotons, at last! Secondary astrophysical
  component and immediate implications for Dark Matter}},  {\em JCAP} {\bf
  1509} (2015), no.~09 023, [\href{http://xxx.lanl.gov/abs/1504.04276}{{\tt
  1504.04276}}].

\bibitem{Lu:2015pta}
B.-Q. Lu and H.-S. Zong, {\it {Limits on dark matter from AMS-02 antiproton and
  positron fraction data}},  {\em Phys. Rev.} {\bf D93} (2016), no.~10 103517,
  [\href{http://xxx.lanl.gov/abs/1510.04032}{{\tt 1510.04032}}].

\bibitem{Cavasonza:2016qem}
L.~A. Cavasonza, H.~Gast, M.~Kramer, M.~Pellen, and S.~Schael, {\it
  {Constraints on leptophilic dark matter from the AMS-02 experiment}},
  \href{http://xxx.lanl.gov/abs/1612.06634}{{\tt 1612.06634}}.

\bibitem{Belotsky:2016tja}
K.~Belotsky, R.~Budaev, A.~Kirillov, and M.~Laletin, {\it {Fermi-LAT kills dark
  matter interpretations of AMS-02 data. Or not?}},
  \href{http://xxx.lanl.gov/abs/1606.01271}{{\tt 1606.01271}}.

\bibitem{Profumo:2016idl}
S.~Profumo, F.~S. Queiroz, and C.~E. Yaguna, {\it {Extending Fermi-LAT and
  H.E.S.S. Limits on Gamma-ray Lines from Dark Matter Annihilation}},  {\em
  Mon. Not. Roy. Astron. Soc.} {\bf 461} (2016), no.~4 3976--3981,
  [\href{http://xxx.lanl.gov/abs/1602.08501}{{\tt 1602.08501}}].

\bibitem{Huang:2016tfo}
X.-J. Huang, C.-C. Wei, Y.-L. Wu, W.-H. Zhang, and Y.-F. Zhou, {\it
  {Antiprotons from dark matter annihilation through light mediators and a
  possible excess in AMS-02 $\bar{p}/p$ data}},
  \href{http://xxx.lanl.gov/abs/1611.01983}{{\tt 1611.01983}}.

\bibitem{Caputo:2016ryl}
R.~Caputo, M.~R. Buckley, P.~Martin, E.~Charles, A.~M. Brooks,
  A.~Drlica-Wagner, J.~M. Gaskins, and M.~Wood, {\it {Search for Gamma-ray
  Emission from Dark Matter Annihilation in the Small Magellanic Cloud with the
  Fermi Large Area Telescope}},  {\em Phys. Rev.} {\bf D93} (2016), no.~6
  062004, [\href{http://xxx.lanl.gov/abs/1603.00965}{{\tt 1603.00965}}].

\bibitem{Jin:2017iwg}
H.-B. Jin, Y.-L. Wu, and Y.-F. Zhou, {\it {Astrophysical background and dark
  matter implication based on latest AMS-02 data}},
  \href{http://xxx.lanl.gov/abs/1701.02213}{{\tt 1701.02213}}.

\bibitem{Mardon:2009rc}
J.~Mardon, Y.~Nomura, D.~Stolarski, and J.~Thaler, {\it {Dark Matter Signals
  from Cascade Annihilations}},  {\em JCAP} {\bf 0905} (2009) 016,
  [\href{http://xxx.lanl.gov/abs/0901.2926}{{\tt 0901.2926}}].

\bibitem{Cerdeno:2015ega}
D.~G. Cerdeno, M.~Peiro, and S.~Robles, {\it {Fits to the Fermi-LAT GeV excess
  with RH sneutrino dark matter: implications for direct and indirect dark
  matter searches and the LHC}},  {\em Phys. Rev.} {\bf D91} (2015), no.~12
  123530, [\href{http://xxx.lanl.gov/abs/1501.01296}{{\tt 1501.01296}}].

\bibitem{Okawa:2016wrr}
S.~Okawa, M.~Tanabashi, and M.~Yamanaka, {\it {Relic Abundance in a Secluded
  Dark Matter Scenario with a Massive Mediator}},  {\em Phys. Rev.} {\bf D95}
  (2017), no.~2 023006, [\href{http://xxx.lanl.gov/abs/1607.08520}{{\tt
  1607.08520}}].

\bibitem{Karwin:2016tsw}
C.~Karwin, S.~Murgia, T.~M.~P. Tait, T.~A. Porter, and P.~Tanedo, {\it {Dark
  Matter Interpretation of the Fermi-LAT Observation Toward the Galactic
  Center}},  {\em Phys. Rev.} {\bf D95} (2017), no.~10 103005,
  [\href{http://xxx.lanl.gov/abs/1612.05687}{{\tt 1612.05687}}].

\bibitem{Kim:2016csm}
Y.~G. Kim, K.~Y. Lee, C.~B. Park, and S.~Shin, {\it {Secluded singlet fermionic
  dark matter driven by the Fermi gamma-ray excess}},  {\em Phys. Rev.} {\bf
  D93} (2016), no.~7 075023, [\href{http://xxx.lanl.gov/abs/1601.05089}{{\tt
  1601.05089}}].

\bibitem{Batell:2017rol}
B.~Batell, T.~Han, and B.~Shams Es~Haghi, {\it {Indirect Detection of Neutrino
  Portal Dark Matter}},  \href{http://xxx.lanl.gov/abs/1704.08708}{{\tt
  1704.08708}}.

\bibitem{Campos:2017odj}
M.~D. Campos, F.~S. Queiroz, C.~E. Yaguna, and C.~Weniger, {\it {Search for
  right-handed neutrinos from dark matter annihilation with gamma-rays}},  {\em
  JCAP} {\bf 1707} (2017), no.~07 016,
  [\href{http://xxx.lanl.gov/abs/1702.06145}{{\tt 1702.06145}}].

\bibitem{Arcadi:2017vis}
G.~Arcadi, F.~S. Queiroz, and C.~Siqueira, {\it {The Semi-Hooperon: Gamma-ray
  and anti-proton excesses in the Galactic Center}},
  \href{http://xxx.lanl.gov/abs/1706.02336}{{\tt 1706.02336}}.

\bibitem{Pospelov:2007mp}
M.~Pospelov, A.~Ritz, and M.~B. Voloshin, {\it {Secluded WIMP Dark Matter}},
  {\em Phys. Lett.} {\bf B662} (2008) 53--61,
  [\href{http://xxx.lanl.gov/abs/0711.4866}{{\tt 0711.4866}}].

\bibitem{Kang:2010mh}
Z.~Kang, T.~Li, T.~Liu, C.~Tong, and J.~M. Yang, {\it {Light Dark Matter from
  the $U(1)_X$ Sector in the NMSSM with Gauge Mediation}},  {\em JCAP} {\bf
  1101} (2011) 028, [\href{http://xxx.lanl.gov/abs/1008.5243}{{\tt
  1008.5243}}].

\bibitem{Murase:2012xs}
K.~Murase and J.~F. Beacom, {\it {Constraining Very Heavy Dark Matter Using
  Diffuse Backgrounds of Neutrinos and Cascaded Gamma Rays}},  {\em JCAP} {\bf
  1210} (2012) 043, [\href{http://xxx.lanl.gov/abs/1206.2595}{{\tt
  1206.2595}}].

\bibitem{Alves:2013tqa}
A.~Alves, S.~Profumo, and F.~S. Queiroz, {\it {The dark $Z^{'}$ portal: direct,
  indirect and collider searches}},  {\em JHEP} {\bf 04} (2014) 063,
  [\href{http://xxx.lanl.gov/abs/1312.5281}{{\tt 1312.5281}}].

\bibitem{Martinez:2014ova}
R.~Mart\'inez, J.~Nisperuza, F.~Ochoa, and J.~P. Rubio, {\it {Scalar dark
  matter with CERN-LEP data and $Z′$ search at the LHC in an $U(1)′$
  model}},  {\em Phys. Rev.} {\bf D90} (2014), no.~9 095004,
  [\href{http://xxx.lanl.gov/abs/1408.5153}{{\tt 1408.5153}}].

\bibitem{DEramo:2016gos}
F.~D'Eramo, B.~J. Kavanagh, and P.~Panci, {\it {You can hide but you have to
  run: direct detection with vector mediators}},  {\em JHEP} {\bf 08} (2016)
  111, [\href{http://xxx.lanl.gov/abs/1605.04917}{{\tt 1605.04917}}].

\bibitem{Celis:2016ayl}
A.~Celis, W.-Z. Feng, and M.~Vollmann, {\it {Dirac dark matter and $b \to s
  \ell^+ \ell^-$ with $\mathrm{U(1)}$ gauge symmetry}},  {\em Phys. Rev.} {\bf
  D95} (2017), no.~3 035018, [\href{http://xxx.lanl.gov/abs/1608.03894}{{\tt
  1608.03894}}].

\bibitem{DeRomeri:2017oxa}
V.~De~Romeri, E.~Fernandez-Martinez, J.~Gehrlein, P.~A.~N. Machado, and
  V.~Niro, {\it {Dark Matter and the elusive $\mathbf{Z'}$ in a dynamical
  Inverse Seesaw scenario}},  \href{http://xxx.lanl.gov/abs/1707.08606}{{\tt
  1707.08606}}.

\bibitem{Ko:2014gha}
P.~Ko, W.-I. Park, and Y.~Tang, {\it {Higgs portal vector dark matter for
  $\mathinner{\mathrm{GeV}}$ scale $\gamma$-ray excess from galactic center}},
  {\em JCAP} {\bf 1409} (2014) 013,
  [\href{http://xxx.lanl.gov/abs/1404.5257}{{\tt 1404.5257}}].

\bibitem{Ardid:2017lry}
M.~Ardid, I.~Felis, A.~Herrero, and J.~A. Mart\'inez-Mora, {\it {Constraining
  Secluded Dark Matter models with the public data from the 79-string IceCube
  search for dark matter in the Sun}},  {\em JCAP} {\bf 1704} (2017), no.~04
  010, [\href{http://xxx.lanl.gov/abs/1701.08863}{{\tt 1701.08863}}].

\bibitem{Cirelli:2016rnw}
M.~Cirelli, P.~Panci, K.~Petraki, F.~Sala, and M.~Taoso, {\it {Dark Matter's
  secret liaisons: phenomenology of a dark U(1) sector with bound states}},
  {\em JCAP} {\bf 1705} (2017), no.~05 036,
  [\href{http://xxx.lanl.gov/abs/1612.07295}{{\tt 1612.07295}}].

\bibitem{Fortes:2015qka}
E.~C. F.~S. Fortes, V.~Pleitez, and F.~W. Stecker, {\it {Secluded WIMPs, dark
  QED with massive photons, and the galactic center gamma-ray excess}},  {\em
  Astropart. Phys.} {\bf 74} (2016) 87--95,
  [\href{http://xxx.lanl.gov/abs/1503.08220}{{\tt 1503.08220}}].

\bibitem{Fortes:2017kca}
E.~C. F.~S. Fortes, V.~Pleitez, and F.~W. Stecker, {\it {Secluded and Putative
  Flipped Dark Matter and Stueckelberg Extensions of the Standard Model}},
  \href{http://xxx.lanl.gov/abs/1703.05275}{{\tt 1703.05275}}.

\bibitem{Pospelov:2008jd}
M.~Pospelov and A.~Ritz, {\it {Astrophysical Signatures of Secluded Dark
  Matter}},  {\em Phys. Lett.} {\bf B671} (2009) 391--397,
  [\href{http://xxx.lanl.gov/abs/0810.1502}{{\tt 0810.1502}}].

\bibitem{Batell:2009zp}
B.~Batell, M.~Pospelov, A.~Ritz, and Y.~Shang, {\it {Solar Gamma Rays Powered
  by Secluded Dark Matter}},  {\em Phys. Rev.} {\bf D81} (2010) 075004,
  [\href{http://xxx.lanl.gov/abs/0910.1567}{{\tt 0910.1567}}].

\bibitem{Abdullah:2014lla}
M.~Abdullah, A.~DiFranzo, A.~Rajaraman, T.~M.~P. Tait, P.~Tanedo, and A.~M.
  Wijangco, {\it {Hidden on-shell mediators for the Galactic Center
  $\gamma$-ray excess}},  {\em Phys. Rev.} {\bf D90} (2014) 035004,
  [\href{http://xxx.lanl.gov/abs/1404.6528}{{\tt 1404.6528}}].

\bibitem{Martin:2014sxa}
A.~Martin, J.~Shelton, and J.~Unwin, {\it {Fitting the Galactic Center
  Gamma-Ray Excess with Cascade Annihilations}},  {\em Phys. Rev.} {\bf D90}
  (2014), no.~10 103513, [\href{http://xxx.lanl.gov/abs/1405.0272}{{\tt
  1405.0272}}].

\bibitem{Dutta:2015ysa}
B.~Dutta, Y.~Gao, T.~Ghosh, and L.~E. Strigari, {\it {Confronting Galactic
  center and dwarf spheroidal gamma-ray observations with cascade annihilation
  models}},  {\em Phys. Rev.} {\bf D92} (2015), no.~7 075019,
  [\href{http://xxx.lanl.gov/abs/1508.05989}{{\tt 1508.05989}}].

\bibitem{Rajaraman:2015xka}
A.~Rajaraman, J.~Smolinsky, and P.~Tanedo, {\it {On-Shell Mediators and
  Top-Charm Dark Matter Models for the Fermi-LAT Galactic Center Excess}},
  \href{http://xxx.lanl.gov/abs/1503.05919}{{\tt 1503.05919}}.

\bibitem{Elor:2015tva}
G.~Elor, N.~L. Rodd, and T.~R. Slatyer, {\it {Multistep cascade annihilations
  of dark matter and the Galactic Center excess}},  {\em Phys. Rev.} {\bf D91}
  (2015) 103531, [\href{http://xxx.lanl.gov/abs/1503.01773}{{\tt 1503.01773}}].

\bibitem{Escudero:2017yia}
M.~Escudero, S.~J. Witte, and D.~Hooper, {\it {Hidden Sector Dark Matter and
  the Galactic Center Gamma-Ray Excess: A Closer Look}},
  \href{http://xxx.lanl.gov/abs/1709.07002}{{\tt 1709.07002}}.

\bibitem{Ackermann:2015zua}
{\bf Fermi-LAT} Collaboration, M.~Ackermann {\em et.~al.}, {\it {Searching for
  Dark Matter Annihilation from Milky Way Dwarf Spheroidal Galaxies with Six
  Years of Fermi Large Area Telescope Data}},  {\em Phys. Rev. Lett.} {\bf 115}
  (2015), no.~23 231301, [\href{http://xxx.lanl.gov/abs/1503.02641}{{\tt
  1503.02641}}].

\bibitem{Abdallah:2016ygi}
{\bf HESS} Collaboration, H.~Abdallah {\em et.~al.}, {\it {Search for dark
  matter annihilations towards the inner Galactic halo from 10 years of
  observations with H.E.S.S}},  {\em Phys. Rev. Lett.} {\bf 117} (2016), no.~11
  111301, [\href{http://xxx.lanl.gov/abs/1607.08142}{{\tt 1607.08142}}].

\bibitem{Slatyer:2015jla}
T.~R. Slatyer, {\it {Indirect dark matter signatures in the cosmic dark ages.
  I. Generalizing the bound on s-wave dark matter annihilation from Planck
  results}},  {\em Phys. Rev.} {\bf D93} (2016), no.~2 023527,
  [\href{http://xxx.lanl.gov/abs/1506.03811}{{\tt 1506.03811}}].

\bibitem{Slatyer:2015kla}
T.~R. Slatyer, {\it {Indirect Dark Matter Signatures in the Cosmic Dark Ages
  II. Ionization, Heating and Photon Production from Arbitrary Energy
  Injections}},  {\em Phys. Rev.} {\bf D93} (2016), no.~2 023521,
  [\href{http://xxx.lanl.gov/abs/1506.03812}{{\tt 1506.03812}}].

\bibitem{Aharonian:2009zk}
{\bf H.E.S.S.} Collaboration, F.~Aharonian, {\it {Spectrum and variability of
  the Galactic Center VHE gamma-ray source HESS J1745-290}},  {\em Astron.
  Astrophys.} {\bf 503} (2009) 817,
  [\href{http://xxx.lanl.gov/abs/0906.1247}{{\tt 0906.1247}}].

\bibitem{Abramowski:2016mir}
{\bf H.E.S.S.} Collaboration, A.~Abramowski {\em et.~al.}, {\it {Acceleration
  of petaelectronvolt protons in the Galactic Centre}},  {\em Nature} {\bf 531}
  (2016) 476, [\href{http://xxx.lanl.gov/abs/1603.07730}{{\tt 1603.07730}}].

\bibitem{Akerib:2015rjg}
{\bf LUX} Collaboration, D.~S. Akerib {\em et.~al.}, {\it {Improved Limits on
  Scattering of Weakly Interacting Massive Particles from Reanalysis of 2013
  LUX Data}},  {\em Phys. Rev. Lett.} {\bf 116} (2016), no.~16 161301,
  [\href{http://xxx.lanl.gov/abs/1512.03506}{{\tt 1512.03506}}].

\bibitem{Amole:2016pye}
{\bf PICO} Collaboration, C.~Amole {\em et.~al.}, {\it {Improved dark matter
  search results from PICO-2L Run 2}},  {\em Phys. Rev.} {\bf D93} (2016),
  no.~6 061101, [\href{http://xxx.lanl.gov/abs/1601.03729}{{\tt 1601.03729}}].

\bibitem{Hehn:2016nll}
{\bf EDELWEISS} Collaboration, L.~Hehn {\em et.~al.}, {\it {Improved
  EDELWEISS-III sensitivity for low-mass WIMPs using a profile likelihood
  approach}},  {\em Eur. Phys. J.} {\bf C76} (2016), no.~10 548,
  [\href{http://xxx.lanl.gov/abs/1607.03367}{{\tt 1607.03367}}].

\bibitem{Aalbers:2016jon}
{\bf DARWIN} Collaboration, J.~Aalbers {\em et.~al.}, {\it {DARWIN: towards the
  ultimate dark matter detector}},  {\em JCAP} {\bf 1611} (2016) 017,
  [\href{http://xxx.lanl.gov/abs/1606.07001}{{\tt 1606.07001}}].

\bibitem{Agnese:2016cpb}
{\bf SuperCDMS} Collaboration, R.~Agnese {\em et.~al.}, {\it {Projected
  Sensitivity of the SuperCDMS SNOLAB experiment}},  {\em Phys. Rev.} {\bf D95}
  (2017), no.~8 082002, [\href{http://xxx.lanl.gov/abs/1610.00006}{{\tt
  1610.00006}}].

\bibitem{Aprile:2016swn}
{\bf XENON100} Collaboration, E.~Aprile {\em et.~al.}, {\it {XENON100 Dark
  Matter Results from a Combination of 477 Live Days}},  {\em Phys. Rev.} {\bf
  D94} (2016), no.~12 122001, [\href{http://xxx.lanl.gov/abs/1609.06154}{{\tt
  1609.06154}}].

\bibitem{Fu:2016ega}
{\bf PandaX-II} Collaboration, C.~Fu {\em et.~al.}, {\it {Spin-Dependent
  Weakly-Interacting-Massive-Particle–Nucleon Cross Section Limits from First
  Data of PandaX-II Experiment}},  {\em Phys. Rev. Lett.} {\bf 118} (2017),
  no.~7 071301, [\href{http://xxx.lanl.gov/abs/1611.06553}{{\tt 1611.06553}}].

\bibitem{Akerib:2016lao}
{\bf LUX} Collaboration, D.~S. Akerib {\em et.~al.}, {\it {Results on the
  Spin-Dependent Scattering of Weakly Interacting Massive Particles on Nucleons
  from the Run 3 Data of the LUX Experiment}},  {\em Phys. Rev. Lett.} {\bf
  116} (2016), no.~16 161302, [\href{http://xxx.lanl.gov/abs/1602.03489}{{\tt
  1602.03489}}].

\bibitem{Cui:2017nnn}
{\bf PandaX-II} Collaboration, X.~Cui {\em et.~al.}, {\it {Dark Matter Results
  From 54-Ton-Day Exposure of PandaX-II Experiment}},
  \href{http://xxx.lanl.gov/abs/1708.06917}{{\tt 1708.06917}}.

\bibitem{Aprile:2017ngb}
{\bf XENON} Collaboration, E.~Aprile {\em et.~al.}, {\it {Search for WIMP
  Inelastic Scattering off Xenon Nuclei with XENON100}},  {\em Phys. Rev.} {\bf
  D96} (2017), no.~2 022008, [\href{http://xxx.lanl.gov/abs/1705.05830}{{\tt
  1705.05830}}].

\bibitem{Aprile:2017yea}
{\bf XENON} Collaboration, E.~Aprile {\em et.~al.}, {\it {Search for Electronic
  Recoil Event Rate Modulation with 4 Years of XENON100 Data}},  {\em Phys.
  Rev. Lett.} {\bf 118} (2017), no.~10 101101,
  [\href{http://xxx.lanl.gov/abs/1701.00769}{{\tt 1701.00769}}].

\bibitem{Fatemighomi:2016ree}
{\bf DEAP-3600} Collaboration, N.~Fatemighomi, {\it {DEAP-3600 dark matter
  experiment}},  in {\em {35th International Symposium on Physics in Collision
  (PIC 2015) Coventry, United Kingdom, September 15-19, 2015}}, 2016.
\newblock \href{http://xxx.lanl.gov/abs/1609.07990}{{\tt 1609.07990}}.

\bibitem{Edsjo:1997bg}
J.~Edsjo and P.~Gondolo, {\it {Neutralino relic density including
  coannihilations}},  {\em Phys. Rev.} {\bf D56} (1997) 1879--1894,
  [\href{http://xxx.lanl.gov/abs/hep-ph/9704361}{{\tt hep-ph/9704361}}].

\bibitem{Duerr:2016tmh}
M.~Duerr, F.~Kahlhoefer, K.~Schmidt-Hoberg, T.~Schwetz, and S.~Vogl, {\it {How
  to save the WIMP: global analysis of a dark matter model with two s-channel
  mediators}},  {\em JHEP} {\bf 09} (2016) 042,
  [\href{http://xxx.lanl.gov/abs/1606.07609}{{\tt 1606.07609}}].

\bibitem{DEramo:2010keq}
F.~D'Eramo and J.~Thaler, {\it {Semi-annihilation of Dark Matter}},  {\em JHEP}
  {\bf 06} (2010) 109, [\href{http://xxx.lanl.gov/abs/1003.5912}{{\tt
  1003.5912}}].

\bibitem{Belanger:2012vp}
G.~Belanger, K.~Kannike, A.~Pukhov, and M.~Raidal, {\it {Impact of
  semi-annihilations on dark matter phenomenology - an example of $Z_N$
  symmetric scalar dark matter}},  {\em JCAP} {\bf 1204} (2012) 010,
  [\href{http://xxx.lanl.gov/abs/1202.2962}{{\tt 1202.2962}}].

\bibitem{Ko:2014loa}
P.~Ko and Y.~Tang, {\it {Galactic center $\gamma$-ray excess in hidden sector
  DM models with dark gauge symmetries: local $Z_{3}$ symmetry as an example}},
   {\em JCAP} {\bf 1501} (2015) 023,
  [\href{http://xxx.lanl.gov/abs/1407.5492}{{\tt 1407.5492}}].

\bibitem{Cai:2015zza}
Y.~Cai and A.~P. Spray, {\it {Fermionic Semi-Annihilating Dark Matter}},  {\em
  JHEP} {\bf 01} (2016) 087, [\href{http://xxx.lanl.gov/abs/1509.08481}{{\tt
  1509.08481}}].

\bibitem{Essig:2009jx}
R.~Essig, N.~Sehgal, and L.~E. Strigari, {\it {Bounds on Cross-sections and
  Lifetimes for Dark Matter Annihilation and Decay into Charged Leptons from
  Gamma-ray Observations of Dwarf Galaxies}},  {\em Phys. Rev.} {\bf D80}
  (2009) 023506, [\href{http://xxx.lanl.gov/abs/0902.4750}{{\tt 0902.4750}}].

\bibitem{Alves:2015pea}
A.~Alves, A.~Berlin, S.~Profumo, and F.~S. Queiroz, {\it {Dark Matter
  Complementarity and the Z$^\prime$ Portal}},  {\em Phys. Rev.} {\bf D92}
  (2015), no.~8 083004, [\href{http://xxx.lanl.gov/abs/1501.03490}{{\tt
  1501.03490}}].

\bibitem{Ducu:2015fda}
O.~Ducu, L.~Heurtier, and J.~Maurer, {\it {LHC signatures of a Z' mediator
  between dark matter and the SU(3) sector}},  {\em JHEP} {\bf 03} (2016) 006,
  [\href{http://xxx.lanl.gov/abs/1509.05615}{{\tt 1509.05615}}].

\bibitem{Alves:2015mua}
A.~Alves, A.~Berlin, S.~Profumo, and F.~S. Queiroz, {\it {Dirac-Fermionic Dark
  Matter in $U(1)_X$ Models}},  \href{http://xxx.lanl.gov/abs/1506.06767}{{\tt
  1506.06767}}.

\bibitem{Okada:2016tci}
N.~Okada and S.~Okada, {\it {$Z^\prime$-portal right-handed neutrino dark
  matter in the minimal U(1)$_X$ extended Standard Model}},
  \href{http://xxx.lanl.gov/abs/1611.02672}{{\tt 1611.02672}}.

\bibitem{Jacques:2016dqz}
T.~Jacques, A.~Katz, E.~Morgante, D.~Racco, M.~Rameez, and A.~Riotto, {\it
  {Complementarity of DM searches in a consistent simplified model: the case of
  $Z′$}},  {\em JHEP} {\bf 10} (2016) 071,
  [\href{http://xxx.lanl.gov/abs/1605.06513}{{\tt 1605.06513}}].

\bibitem{Karam:2015jta}
A.~Karam and K.~Tamvakis, {\it {Dark matter and neutrino masses from a
  scale-invariant multi-Higgs portal}},  {\em Phys. Rev.} {\bf D92} (2015),
  no.~7 075010, [\href{http://xxx.lanl.gov/abs/1508.03031}{{\tt 1508.03031}}].

\bibitem{Karam:2016rsz}
A.~Karam and K.~Tamvakis, {\it {Dark Matter from a Classically Scale-Invariant
  $SU(3)_X$}},  {\em Phys. Rev.} {\bf D94} (2016), no.~5 055004,
  [\href{http://xxx.lanl.gov/abs/1607.01001}{{\tt 1607.01001}}].

\bibitem{Arcadi:2016kmk}
G.~Arcadi, C.~Gross, O.~Lebedev, Y.~Mambrini, S.~Pokorski, and T.~Toma, {\it
  {Multicomponent Dark Matter from Gauge Symmetry}},  {\em JHEP} {\bf 12}
  (2016) 081, [\href{http://xxx.lanl.gov/abs/1611.00365}{{\tt 1611.00365}}].

\bibitem{Strassler:2006im}
M.~J. Strassler and K.~M. Zurek, {\it {Echoes of a hidden valley at hadron
  colliders}},  {\em Phys. Lett.} {\bf B651} (2007) 374--379,
  [\href{http://xxx.lanl.gov/abs/hep-ph/0604261}{{\tt hep-ph/0604261}}].

\bibitem{Cassel:2009pu}
S.~Cassel, D.~M. Ghilencea, and G.~G. Ross, {\it {Electroweak and Dark Matter
  Constraints on a Z-prime in Models with a Hidden Valley}},  {\em Nucl. Phys.}
  {\bf B827} (2010) 256--280, [\href{http://xxx.lanl.gov/abs/0903.1118}{{\tt
  0903.1118}}].

\bibitem{Arcadi:2017jqd}
G.~Arcadi, P.~Ghosh, Y.~Mambrini, M.~Pierre, and F.~S. Queiroz, {\it {$Z'$
  portal to Chern-Simons Dark Matter}},
  \href{http://xxx.lanl.gov/abs/1706.04198}{{\tt 1706.04198}}.

\bibitem{Navarro:2008kc}
J.~F. Navarro, A.~Ludlow, V.~Springel, J.~Wang, M.~Vogelsberger, S.~D.~M.
  White, A.~Jenkins, C.~S. Frenk, and A.~Helmi, {\it {The Diversity and
  Similarity of Cold Dark Matter Halos}},  {\em Mon. Not. Roy. Astron. Soc.}
  {\bf 402} (2010) 21, [\href{http://xxx.lanl.gov/abs/0810.1522}{{\tt
  0810.1522}}].

\bibitem{Graham:2005xx}
A.~W. Graham, D.~Merritt, B.~Moore, J.~Diemand, and B.~Terzic, {\it {Empirical
  models for Dark Matter Halos. I. Nonparametric Construction of Density
  Profiles and Comparison with Parametric Models}},  {\em Astron. J.} {\bf 132}
  (2006) 2685--2700, [\href{http://xxx.lanl.gov/abs/astro-ph/0509417}{{\tt
  astro-ph/0509417}}].

\bibitem{Burkert:1995yz}
A.~Burkert, {\it {The Structure of dark matter halos in dwarf galaxies}},  {\em
  IAU Symp.} {\bf 171} (1996) 175,
  [\href{http://xxx.lanl.gov/abs/astro-ph/9504041}{{\tt astro-ph/9504041}}].
  [Astrophys. J.447,L25(1995)].

\bibitem{Salucci:2000ps}
P.~Salucci and A.~Burkert, {\it {Dark matter scaling relations}},  {\em
  Astrophys. J.} {\bf 537} (2000) L9--L12,
  [\href{http://xxx.lanl.gov/abs/astro-ph/0004397}{{\tt astro-ph/0004397}}].

\bibitem{Cirelli:2010xx}
M.~Cirelli, G.~Corcella, A.~Hektor, G.~Hutsi, M.~Kadastik, P.~Panci, M.~Raidal,
  F.~Sala, and A.~Strumia, {\it {PPPC 4 DM ID: A Poor Particle Physicist
  Cookbook for Dark Matter Indirect Detection}},  {\em JCAP} {\bf 1103} (2011)
  051, [\href{http://xxx.lanl.gov/abs/1012.4515}{{\tt 1012.4515}}]. [Erratum:
  JCAP1210,E01(2012)].

\bibitem{Sjostrand:2014zea}
T.~Sjostrand, S.~Ask, J.~R. Christiansen, R.~Corke, N.~Desai, P.~Ilten,
  S.~Mrenna, S.~Prestel, C.~O. Rasmussen, and P.~Z. Skands, {\it {An
  Introduction to PYTHIA 8.2}},  {\em Comput. Phys. Commun.} {\bf 191} (2015)
  159--177, [\href{http://xxx.lanl.gov/abs/1410.3012}{{\tt 1410.3012}}].

\bibitem{Padmanabhan:2005es}
N.~Padmanabhan and D.~P. Finkbeiner, {\it {Detecting dark matter annihilation
  with CMB polarization: Signatures and experimental prospects}},  {\em Phys.
  Rev.} {\bf D72} (2005) 023508,
  [\href{http://xxx.lanl.gov/abs/astro-ph/0503486}{{\tt astro-ph/0503486}}].

\bibitem{Gehrels:1999ri}
N.~Gehrels and P.~Michelson, {\it {GLAST: The next generation high-energy
  gamma-ray astronomy mission}},  {\em Astropart. Phys.} {\bf 11} (1999)
  277--282.

\bibitem{Abdo:2010ex}
{\bf Fermi-LAT} Collaboration, A.~A. Abdo {\em et.~al.}, {\it {Observations of
  Milky Way Dwarf Spheroidal galaxies with the Fermi-LAT detector and
  constraints on Dark Matter models}},  {\em Astrophys. J.} {\bf 712} (2010)
  147--158, [\href{http://xxx.lanl.gov/abs/1001.4531}{{\tt 1001.4531}}].

\bibitem{Ackermann:2011wa}
{\bf Fermi-LAT collaboration} Collaboration, M.~Ackermann {\em et.~al.}, {\it
  {Constraining Dark Matter Models from a Combined Analysis of Milky Way
  Satellites with the Fermi Large Area Telescope}},  {\em Phys.Rev.Lett.} {\bf
  107} (2011) 241302, [\href{http://xxx.lanl.gov/abs/1108.3546}{{\tt
  1108.3546}}].

\bibitem{Ackermann:2012nb}
{\bf Fermi-LAT} Collaboration, M.~Ackermann {\em et.~al.}, {\it {Search for
  Dark Matter Satellites using the FERMI-LAT}},  {\em Astrophys. J.} {\bf 747}
  (2012) 121, [\href{http://xxx.lanl.gov/abs/1201.2691}{{\tt 1201.2691}}].

\bibitem{Ackermann:2013yva}
{\bf Fermi-LAT Collaboration} Collaboration, M.~Ackermann {\em et.~al.}, {\it
  {Dark Matter Constraints from Observations of 25 Milky Way Satellite Galaxies
  with the Fermi Large Area Telescope}},  {\em Phys.Rev.} {\bf D89} (2014)
  042001, [\href{http://xxx.lanl.gov/abs/1310.0828}{{\tt 1310.0828}}].

\bibitem{Drlica-Wagner:2015xua}
{\bf DES, Fermi-LAT} Collaboration, A.~Drlica-Wagner {\em et.~al.}, {\it
  {Search for Gamma-Ray Emission from DES Dwarf Spheroidal Galaxy Candidates
  with Fermi-LAT Data}},  {\em Astrophys. J.} {\bf 809} (2015), no.~1 L4,
  [\href{http://xxx.lanl.gov/abs/1503.02632}{{\tt 1503.02632}}].

\bibitem{Rico:2015nya}
{\bf Fermi-LAT, MAGIC} Collaboration, J.~Rico, M.~Wood, A.~Drlica-Wagner, and
  J.~Aleksic, {\it {Limits to dark matter properties from a combined analysis
  of MAGIC and $Fermi$-LAT observations of dwarf satellite galaxies}},  {\em
  PoS} {\bf ICRC2015} (2016) 1206,
  [\href{http://xxx.lanl.gov/abs/1508.05827}{{\tt 1508.05827}}].

\bibitem{Ahnen:2016qkx}
{\bf Fermi-LAT, MAGIC} Collaboration, M.~L. Ahnen {\em et.~al.}, {\it {Limits
  to dark matter annihilation cross-section from a combined analysis of MAGIC
  and Fermi-LAT observations of dwarf satellite galaxies}},  {\em JCAP} {\bf
  1602} (2016), no.~02 039, [\href{http://xxx.lanl.gov/abs/1601.06590}{{\tt
  1601.06590}}].

\bibitem{Blumenthal:1984bp}
G.~R. Blumenthal, S.~M. Faber, J.~R. Primack, and M.~J. Rees, {\it {Formation
  of Galaxies and Large Scale Structure with Cold Dark Matter}},  {\em Nature}
  {\bf 311} (1984) 517--525.

\bibitem{Peebles:1984zz}
P.~J.~E. Peebles, {\it {Dark matter and the origin of galaxies and globular
  star clusters}},  {\em Astrophys. J.} {\bf 277} (1984) 470--477.

\bibitem{Dekek:1986gu}
A.~Dekel and J.~Silk, {\it {The origin of dwarf galaxies, cold dark matter, and
  biased galaxy formation}},  {\em Astrophys. J.} {\bf 303} (1986) 39--55.

\bibitem{Kent:1987zz}
S.~M. Kent, {\it {Dark matter in spiral galaxies. II - Galaxies with H I
  rotation curves}},  {\em Astron. J.} {\bf 93} (1987) 816--832.

\bibitem{Moore:1995pb}
B.~Moore, {\it {Constraints on the global mass to light ratios and extent of
  dark matter halos in globular clusters and dwarf spheroidals}},  {\em
  Astrophys. J.} {\bf 461} (1996) L13,
  [\href{http://xxx.lanl.gov/abs/astro-ph/9511147}{{\tt astro-ph/9511147}}].

\bibitem{Ferrara:1999ry}
A.~Ferrara and E.~Tolstoy, {\it {The role of stellar feedback and dark matter
  in the evolution of dwarf galaxies}},  {\em Mon. Not. Roy. Astron. Soc.} {\bf
  313} (2000) 291, [\href{http://xxx.lanl.gov/abs/astro-ph/9905280}{{\tt
  astro-ph/9905280}}].

\bibitem{Firmani:2000qe}
C.~Firmani, E.~D'Onghia, G.~Chincarini, X.~Hernandez, and V.~Avila-Reese, {\it
  {Constraints on dark matter physics from dwarf galaxies through galaxy
  cluster haloes}},  {\em Mon. Not. Roy. Astron. Soc.} {\bf 321} (2001) 713,
  [\href{http://xxx.lanl.gov/abs/astro-ph/0005001}{{\tt astro-ph/0005001}}].

\bibitem{vandenBosch:2000rza}
F.~C. van~den Bosch and R.~A. Swaters, {\it {Dwarf galaxy rotation curves and
  the core problem of dark matter halos}},  {\em Mon. Not. Roy. Astron. Soc.}
  {\bf 325} (2001) 1017, [\href{http://xxx.lanl.gov/abs/astro-ph/0006048}{{\tt
  astro-ph/0006048}}].

\bibitem{Walker:2009zp}
M.~G. Walker, M.~Mateo, E.~W. Olszewski, J.~Penarrubia, N.~W. Evans, and
  G.~Gilmore, {\it {A Universal Mass Profile for Dwarf Spheroidal Galaxies}},
  {\em Astrophys. J.} {\bf 704} (2009) 1274--1287,
  [\href{http://xxx.lanl.gov/abs/0906.0341}{{\tt 0906.0341}}]. [Erratum:
  Astrophys. J.710,886(2010)].

\bibitem{Pasetto:2010se}
S.~Pasetto, E.~K. Grebel, P.~Berczik, R.~Spurzem, and W.~Dehnen, {\it {On the
  isolated dwarf galaxies: from cuspy to flat dark matter density profiles and
  metallicity gradients}},  {\em Astron. Astrophys.} {\bf 514} (2010) A47,
  [\href{http://xxx.lanl.gov/abs/1002.1085}{{\tt 1002.1085}}].

\bibitem{DelPopolo:2011cj}
A.~Del~Popolo, {\it {Density profile slope in Dwarfs and environment}},  {\em
  Mon. Not. Roy. Astron. Soc.} {\bf 419} (2012) 971--984,
  [\href{http://xxx.lanl.gov/abs/1105.0090}{{\tt 1105.0090}}].

\bibitem{Charbonnier:2011ft}
A.~Charbonnier {\em et.~al.}, {\it {Dark matter profiles and annihilation in
  dwarf spheroidal galaxies: prospectives for present and future gamma-ray
  observatories - I. The classical dSphs}},  {\em Mon. Not. Roy. Astron. Soc.}
  {\bf 418} (2011) 1526--1556, [\href{http://xxx.lanl.gov/abs/1104.0412}{{\tt
  1104.0412}}].

\bibitem{Jardel:2012am}
J.~R. Jardel, K.~Gebhardt, M.~H. Fabricius, N.~Drory, and M.~J. Williams, {\it
  {Measuring Dark Matter Profiles Non-Parametrically in Dwarf Spheroidals: An
  Application to Draco}},  {\em Astrophys. J.} {\bf 763} (2013) 91,
  [\href{http://xxx.lanl.gov/abs/1211.5376}{{\tt 1211.5376}}].

\bibitem{Collins:2013eek}
M.~L.~M. Collins {\em et.~al.}, {\it {The masses of Local Group dwarf
  spheroidal galaxies: The death of the universal mass profile}},  {\em
  Astrophys. J.} {\bf 783} (2014) 7,
  [\href{http://xxx.lanl.gov/abs/1309.3053}{{\tt 1309.3053}}].

\bibitem{Laporte:2013fwa}
C.~F.~P. Laporte, M.~G. Walker, and J.~Penarrubia, {\it {Measuring the slopes
  of mass profiles for dwarf spheroidals in triaxial CDM potentials}},  {\em
  Mon. Not. Roy. Astron. Soc.} {\bf 433} (2013) 54,
  [\href{http://xxx.lanl.gov/abs/1303.1534}{{\tt 1303.1534}}].

\bibitem{Geringer-Sameth:2014yza}
A.~Geringer-Sameth, S.~M. Koushiappas, and M.~Walker, {\it {Dwarf galaxy
  annihilation and decay emission profiles for dark matter experiments}},  {\em
  Astrophys. J.} {\bf 801} (2015), no.~2 74,
  [\href{http://xxx.lanl.gov/abs/1408.0002}{{\tt 1408.0002}}].

\bibitem{Adams:2014bda}
J.~J. Adams {\em et.~al.}, {\it {Dwarf Galaxy Dark Matter Density Profiles
  Inferred from Stellar and Gas Kinematics}},  {\em Astrophys. J.} {\bf 789}
  (2014), no.~1 63, [\href{http://xxx.lanl.gov/abs/1405.4854}{{\tt
  1405.4854}}].

\bibitem{Bonnivard:2015tta}
V.~Bonnivard, C.~Combet, D.~Maurin, A.~Geringer-Sameth, S.~M. Koushiappas,
  M.~G. Walker, M.~Mateo, E.~W. Olszewski, and J.~I. Bailey~III, {\it {Dark
  matter annihilation and decay profiles for the Reticulum II dwarf spheroidal
  galaxy}},  {\em Astrophys. J.} {\bf 808} (2015), no.~2 L36,
  [\href{http://xxx.lanl.gov/abs/1504.03309}{{\tt 1504.03309}}].

\bibitem{Chiappo:2016xfs}
A.~Chiappo, J.~Cohen-Tanugi, J.~Conrad, L.~E. Strigari, B.~Anderson, and M.~A.
  Sanchez-Conde, {\it {Dwarf spheroidal J-factors without priors: A
  likelihood-based analysis for indirect dark matter searches}},  {\em Mon.
  Not. Roy. Astron. Soc.} {\bf 466} (2017), no.~1 669--676,
  [\href{http://xxx.lanl.gov/abs/1608.07111}{{\tt 1608.07111}}].

\bibitem{Queiroz:2016zwd}
F.~S. Queiroz, C.~E. Yaguna, and C.~Weniger, {\it {Gamma-ray Limits on Neutrino
  Lines}},  {\em JCAP} {\bf 1605} (2016), no.~05 050,
  [\href{http://xxx.lanl.gov/abs/1602.05966}{{\tt 1602.05966}}].

\bibitem{Abramowski:2011hc}
{\bf HESS} Collaboration, A.~Abramowski {\em et.~al.}, {\it {Search for a Dark
  Matter annihilation signal from the Galactic Center halo with H.E.S.S}},
  {\em Phys. Rev. Lett.} {\bf 106} (2011) 161301,
  [\href{http://xxx.lanl.gov/abs/1103.3266}{{\tt 1103.3266}}].

\bibitem{Lefranc:2015vza}
{\bf HESS} Collaboration, V.~Lefranc and E.~Moulin, {\it {Dark matter search in
  the inner Galactic halo with H.E.S.S. I and H.E.S.S. II}},  in {\em
  {Proceedings, 34th International Cosmic Ray Conference (ICRC 2015)}}, 2015.
\newblock \href{http://xxx.lanl.gov/abs/1509.04123}{{\tt 1509.04123}}.

\bibitem{Ade:2013zuv}
{\bf Planck} Collaboration, P.~A.~R. Ade {\em et.~al.}, {\it {Planck 2013
  results. XVI. Cosmological parameters}},  {\em Astron. Astrophys.} {\bf 571}
  (2014) A16, [\href{http://xxx.lanl.gov/abs/1303.5076}{{\tt 1303.5076}}].

\bibitem{Planck:2013nga}
{\bf Planck} Collaboration, P.~A.~R. Ade {\em et.~al.}, {\it {Planck
  intermediate results. XVI. Profile likelihoods for cosmological parameters}},
   {\em Astron. Astrophys.} {\bf 566} (2014) A54,
  [\href{http://xxx.lanl.gov/abs/1311.1657}{{\tt 1311.1657}}].

\bibitem{Ade:2015xua}
{\bf Planck} Collaboration, P.~A.~R. Ade {\em et.~al.}, {\it {Planck 2015
  results. XIII. Cosmological parameters}},  {\em Astron. Astrophys.} {\bf 594}
  (2016) A13, [\href{http://xxx.lanl.gov/abs/1502.01589}{{\tt 1502.01589}}].

\bibitem{Galli:2009zc}
S.~Galli, F.~Iocco, G.~Bertone, and A.~Melchiorri, {\it {CMB constraints on
  Dark Matter models with large annihilation cross-section}},  {\em Phys. Rev.}
  {\bf D80} (2009) 023505, [\href{http://xxx.lanl.gov/abs/0905.0003}{{\tt
  0905.0003}}].

\bibitem{Galli:2011rz}
S.~Galli, F.~Iocco, G.~Bertone, and A.~Melchiorri, {\it {Updated CMB
  constraints on Dark Matter annihilation cross-sections}},  {\em Phys. Rev.}
  {\bf D84} (2011) 027302, [\href{http://xxx.lanl.gov/abs/1106.1528}{{\tt
  1106.1528}}].

\bibitem{Finkbeiner:2011dx}
D.~P. Finkbeiner, S.~Galli, T.~Lin, and T.~R. Slatyer, {\it {Searching for Dark
  Matter in the CMB: A Compact Parameterization of Energy Injection from New
  Physics}},  {\em Phys. Rev.} {\bf D85} (2012) 043522,
  [\href{http://xxx.lanl.gov/abs/1109.6322}{{\tt 1109.6322}}].

\bibitem{Madhavacheril:2013cna}
M.~S. Madhavacheril, N.~Sehgal, and T.~R. Slatyer, {\it {Current Dark Matter
  Annihilation Constraints from CMB and Low-Redshift Data}},  {\em Phys. Rev.}
  {\bf D89} (2014) 103508, [\href{http://xxx.lanl.gov/abs/1310.3815}{{\tt
  1310.3815}}].

\bibitem{Lopez-Honorez:2013lcm}
L.~Lopez-Honorez, O.~Mena, S.~Palomares-Ruiz, and A.~C. Vincent, {\it
  {Constraints on dark matter annihilation from CMB observationsbefore
  Planck}},  {\em JCAP} {\bf 1307} (2013) 046,
  [\href{http://xxx.lanl.gov/abs/1303.5094}{{\tt 1303.5094}}].

\bibitem{Poulin:2015pna}
V.~Poulin, P.~D. Serpico, and J.~Lesgourgues, {\it {Dark Matter annihilations
  in halos and high-redshift sources of reionization of the universe}},  {\em
  JCAP} {\bf 1512} (2015), no.~12 041,
  [\href{http://xxx.lanl.gov/abs/1508.01370}{{\tt 1508.01370}}].

\bibitem{Shu:2007wg}
J.~Shu, {\it {Unitarity Bounds for New Physics from Axial Coupling at LHC}},
  {\em Phys. Rev.} {\bf D78} (2008) 096004,
  [\href{http://xxx.lanl.gov/abs/0711.2516}{{\tt 0711.2516}}].

\bibitem{Kahlhoefer:2015bea}
F.~Kahlhoefer, K.~Schmidt-Hoberg, T.~Schwetz, and S.~Vogl, {\it {Implications
  of unitarity and gauge invariance for simplified dark matter models}},  {\em
  JHEP} {\bf 02} (2016) 016, [\href{http://xxx.lanl.gov/abs/1510.02110}{{\tt
  1510.02110}}].

\bibitem{Robertson:2009bh}
B.~Robertson and A.~Zentner, {\it {Dark Matter Annihilation Rates with
  Velocity-Dependent Annihilation Cross Sections}},  {\em Phys. Rev.} {\bf D79}
  (2009) 083525, [\href{http://xxx.lanl.gov/abs/0902.0362}{{\tt 0902.0362}}].

\bibitem{Campbell:2010xc}
S.~Campbell, B.~Dutta, and E.~Komatsu, {\it {Effects of Velocity-Dependent Dark
  Matter Annihilation on the Energy Spectrum of the Extragalactic Gamma-ray
  Background}},  {\em Phys. Rev.} {\bf D82} (2010) 095007,
  [\href{http://xxx.lanl.gov/abs/1009.3530}{{\tt 1009.3530}}].

\bibitem{Hisano:2011dc}
J.~Hisano, M.~Kawasaki, K.~Kohri, T.~Moroi, K.~Nakayama, and T.~Sekiguchi, {\it
  {Cosmological constraints on dark matter models with velocity-dependent
  annihilation cross section}},  {\em Phys. Rev.} {\bf D83} (2011) 123511,
  [\href{http://xxx.lanl.gov/abs/1102.4658}{{\tt 1102.4658}}].

\bibitem{Arina:2014yna}
C.~Arina, E.~Del~Nobile, and P.~Panci, {\it {Dark Matter with
  Pseudoscalar-Mediated Interactions Explains the DAMA Signal and the Galactic
  Center Excess}},  {\em Phys. Rev. Lett.} {\bf 114} (2015) 011301,
  [\href{http://xxx.lanl.gov/abs/1406.5542}{{\tt 1406.5542}}].

\bibitem{Zhao:2016xie}
Y.~Zhao, X.-J. Bi, H.-Y. Jia, P.-F. Yin, and F.-R. Zhu, {\it {Constraint on the
  velocity dependent dark matter annihilation cross section from Fermi-LAT
  observations of dwarf galaxies}},  {\em Phys. Rev.} {\bf D93} (2016), no.~8
  083513, [\href{http://xxx.lanl.gov/abs/1601.02181}{{\tt 1601.02181}}].

\bibitem{Goncalves:2016iyg}
D.~Goncalves, P.~A.~N. Machado, and J.~M. No, {\it {Simplified Models for Dark
  Matter Face their Consistent Completions}},  {\em Phys. Rev.} {\bf D95}
  (2017), no.~5 055027, [\href{http://xxx.lanl.gov/abs/1611.04593}{{\tt
  1611.04593}}].

\bibitem{Gonzalez-Morales:2017jkx}
A.~X. Gonzalez-Morales, S.~Profumo, and J.~Reynoso-C\'ordova, {\it {Prospects
  for indirect MeV Dark Matter detection with Gamma Rays in light of Cosmic
  Microwave Background Constraints}},  {\em Phys. Rev.} {\bf D96} (2017), no.~6
  063520, [\href{http://xxx.lanl.gov/abs/1705.00777}{{\tt 1705.00777}}].

\bibitem{Belikov:2012ty}
A.~V. Belikov, G.~Zaharijas, and J.~Silk, {\it {Study of the Gamma-ray Spectrum
  from the Galactic Center in view of Multi-TeV Dark Matter Candidates}},  {\em
  Phys. Rev.} {\bf D86} (2012) 083516,
  [\href{http://xxx.lanl.gov/abs/1207.2412}{{\tt 1207.2412}}].

\bibitem{Belikov:2013laa}
A.~V. Belikov and J.~Silk, {\it {Superexponential Cutoff as a Probe of
  Annihilating Dark Matter}},  {\em Phys. Rev. Lett.} {\bf 111} (2013), no.~7
  071302, [\href{http://xxx.lanl.gov/abs/1304.5238}{{\tt 1304.5238}}].

\bibitem{Belikov:2016fwv}
A.~V. Belikov, E.~Moulin, and J.~Silk, {\it {Study of the very high energy
  gamma-ray spectrum from the Galactic Center and future prospects}},  {\em
  Phys. Rev.} {\bf D94} (2016), no.~10 103005,
  [\href{http://xxx.lanl.gov/abs/1610.10003}{{\tt 1610.10003}}].

\end{thebibliography}\endgroup

\end{document}